\documentclass[conference]{IEEEtran}
\IEEEoverridecommandlockouts
\usepackage{lineno,hyperref}
\usepackage{graphicx}
\usepackage{epstopdf}
\usepackage{algorithm}
\usepackage{algorithmicx}
\usepackage{algpseudocode}
\usepackage{amsmath}
\usepackage{amstext}
\usepackage{amssymb}
\usepackage{caption}
\usepackage{subfigure}
\usepackage{color}
\usepackage{multirow}
\usepackage{subfigure}
\usepackage{listings}
\usepackage{xcolor}
\begin{document}

\title{BcMON: Blockchain Middleware for Offline Networks}

\maketitle

\begin{abstract}

Blockchain is becoming a new generation of information infrastructures. However, the current blockchain solutions rely on a continuous connectivity network to query and modify the state of the blockchain. The emerging satellite technology seems to be a good catalyst to forward offline transactions to the blockchain. However, this approach suffers expensive costs, difficult interoperability, and limited computation problems. Therefore, we propose BcMON, the first blockchain middleware for offline networks. BcMON incorporates three innovative designs: 1) it reduces the costs of offline transactions accessing the blockchain through Short Message Service (SMS), 2) it validates the authenticity of offline cross-chain transactions by two-phase consensus, 3) it supports offline clients to perform complex queries and computations on the blockchains. The prototype of BcMON has been implemented to evaluate the performance of the proposed middleware, which can show its stability, efficiency, and scalability.

\end{abstract}


\section{Introduction}

Blockchain was first used in a peer-to-peer electronic cash system \cite{nakamoto2008bitcoin} to provide immutability through the chain data structure, consensus, and redundant storage. Clients can query and modify the state of the blockchain by full and light nodes. A full node has a complete blockchain ledger. It exploits memory to synchronize blockchain data, broadcast transactions, verify transactions and update data in real-time. A light node maintains the block headers of the ledger. It can validate transactions by interacting with full nodes. Since blockchain is an application protocol built on the Internet, clients rely on a continuous connectivity network to interact with blockchain nodes \cite{cong2021dtnb}.

However, 2.9 billion people in the world have never used the Internet, and parts of people who have used the Internet have only occasional access to the Internet \cite{ituit}. Although blockchain technologies have developed rapidly in recent years, \textit{how to conduct trusted interactions between the blockchain and offline clients becomes a key challenge}.

The feasible solution exploits satellites to relay offline transactions to blockchain nodes \cite{ blockstream}\cite{spacechain}. Clients buy specific transmission devices to connect to satellites and initiate transactions. However, this approach has the following problems. \textit{First, Expensive Costs.} It requires blockchain solutions to cooperate with centralized satellite companies to provide access services. And the costs of connections are passed on to clients. Moreover, it requires clients purchase expensive launch equipment to use the service. \textit{Second, Difficult Interoperability.} \cite{blockstream} and \cite{smartsmesh} build offline networks for Bitcoin and Ethereum respectively. It is difficult to adapt to different blockchains quickly, and it is also impossible to achieve interoperability between them. \textit{Third, Limited Computation.} Bitcoin's blockchain alone is 385GB as of March 2022 \cite{statista}. On-chain data is becoming a valuable asset. Currently, offline clients can not use on-chain data to perform complicated computations. 

To deal with the challenges above, we propose BcMON, the blockchain middleware for offline networks. BcMON is composed of three components, including Offline Blockchain Services (OFBS), Cross-chain Blockchain Service (CCBS), and Computing Blockchain Service (CPBS). OFBS realizes the interaction between offline clients and the blockchain. Based on OFBS, CCBS realizes the interoperability between offline clients and multiple blockchains. Based on OFBS and CCBS, CPBS realizes that offline clients perform complex queries and computations on the blockchains.

Main contributions of this paper are summarized as follows.

\begin{itemize}
  \item We propose a blockchain middleware for the offline network. To the best of our knowledge, this is the first work on blockchain middleware for the offline network. We believe this is a timely study, as the blockchain is widely used in many scenarios.
    
  \item We design SMS-based OFBS to reduce the costs of offline transactions accessing the blockchain. To protect the integrity of offline transactions, we propose a reliable interaction mechanism based on offline channels.
      
  \item We propose CCBS to validate the authenticity of offline cross-chain transactions. Moreover, we propose a two-phase consensus to protect the atomicity and integrity of offline cross-chain transactions.
    
  \item We propose CPBS to support offline clients to perform complex queries and computations on the blockchains. And we design a threshold signature-based interaction mechanism to help offline clients validate results.
  
\end{itemize}

The rest of the paper is organized as follows. Offline blockchain service design is introduced in Section \uppercase\expandafter{\romannumeral2}. Cross-chain blockchain service design is presented in Section \uppercase\expandafter{\romannumeral3}. Computing blockchain service design is elaborated in Section \uppercase\expandafter{\romannumeral4}. The proposed method is evaluated in Section \uppercase\expandafter{\romannumeral5}. Related work is discussed in Section \uppercase\expandafter{\romannumeral6} and and the paper is concluded in Section \uppercase\expandafter{\romannumeral7}.

\section{Offline Blockchain Service Design}
\label{ofbs-section}

\begin{figure}[!t]
  \centering
  \includegraphics[width=3.5in]{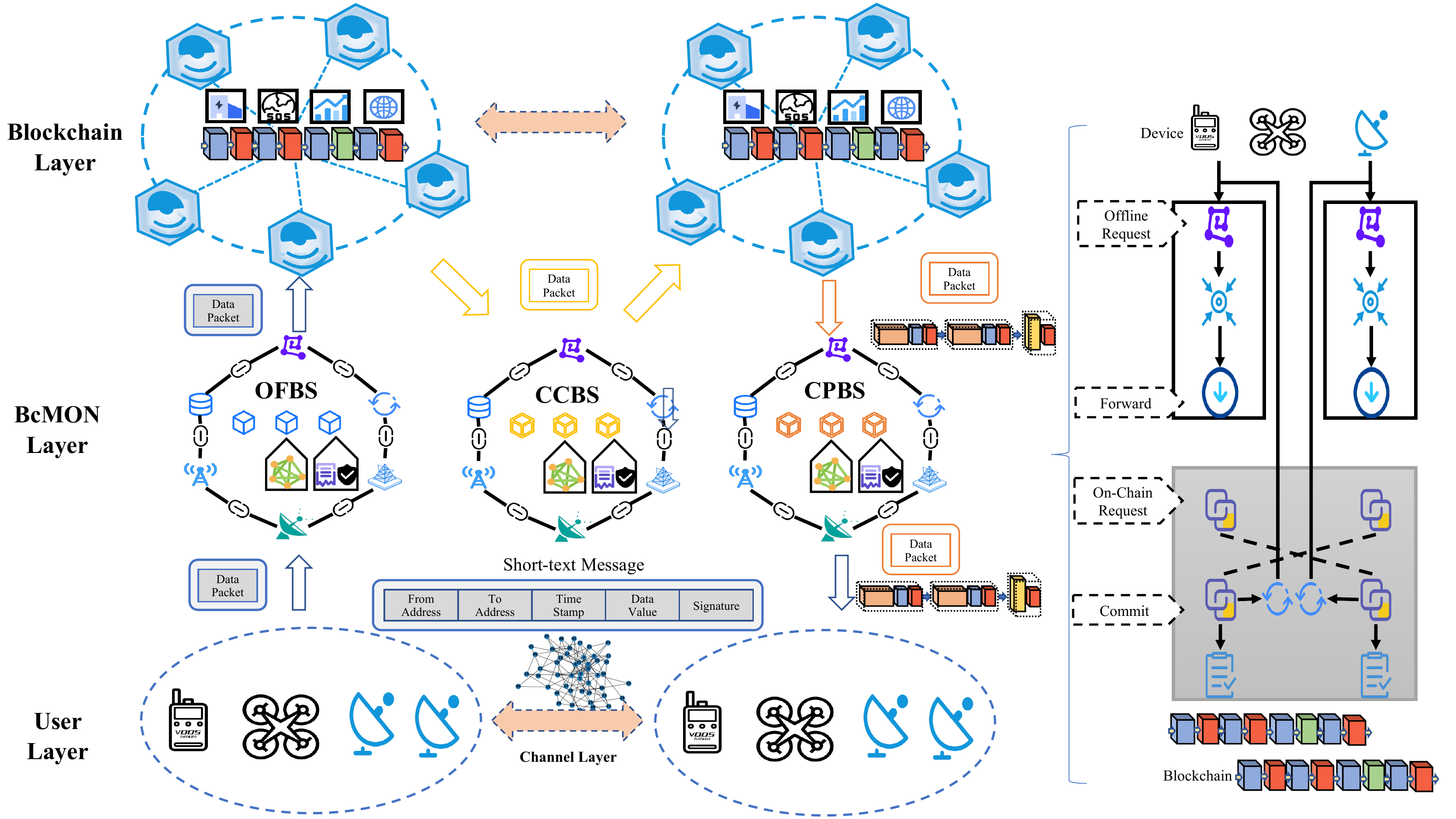}
      \caption{Architecture of BcMON}
      \label{architecture}
\end{figure}

This section mainly describes the architecture of BcMON, as shown in Figure \ref{architecture}. 

\subsection{Architecture}

The architecture is divided into four layers: the user layer, channel layer, BcMON layer, and blockchain layer. And the BcMON layer includes OFBS, CCBS, and CPBS. 

\paragraph{User Layer} The user layer is composed of offline clients in a weak communication environment. Offline clients send data packets containing transactions to the BcMON layer, which is the access to the blockchain network.

Offline clients are required to own accounts in the blockchain. Then they can exploit services supported by the blockchain middleware. Some offline clients who have occasional opportunities to connect to the Internet can independently generate public keys, private keys, and account addresses. Other offline clients who have no chance to connect to the Internet can entrust a trusted third party to escrow the public and private keys, like operators of mobile phones.

\paragraph{BcMON Layer} The BcMON layer consists of three components, including OFBS, CCBS, and CPBS. The underlying network architecture of the three components is a peer-to-peer network. And nodes of the three components are composed of infrastructures with strong computing and storage capabilities, like cell towers. 

OFBS receives SMS messages from offline clients and forwards them to specific middleware according to demands. If offline clients initiate transactions with a single blockchain, OFBS forwards data packets to the specific blockchains. See details in Section \ref{ofbs_workflow}. If offline clients initiate transactions with multiple blockchains, OFBS forwards data packets to CCBS, and CCBS will perform operations, see details in Section \ref{ccbs_workflow}. If offline clients initiate a transaction with complex computation on on-chain data, OFBS forwards data packets to CPBS, and CPBS will perform operations, see details in Section \ref{cpbs_workflow}. 

\paragraph{Channel Layer} The channel layer consists of virtual links between offline clients. Since offline clients have limited resources and some blockchains have limited throughput, frequent interactions with blockchains via OFBS are inefficient and wasteful of energy. Therefore, OFBS sets up virtual links between offline clients, supported by smart contracts and SMS services. Transactions initiated in the virtual links are not immediately submitted to the blockchain until the number of transactions is up to the max buffer size or meets other conditions.

\paragraph{Blockchain Layer} The blockchain layer consists of multiple blockchains. Each blockchain is an independent decentralized network. The way that blockchains interact with offline clients is OFBS. CCBS provides blockchain cross-chain interoperability. CPBS provides complex queries and computing power.

\subsection{Workflow}
\label{ofbs_workflow}

\begin{figure}[!t]
  \centering
  \includegraphics[width=3.5in, height=4in]{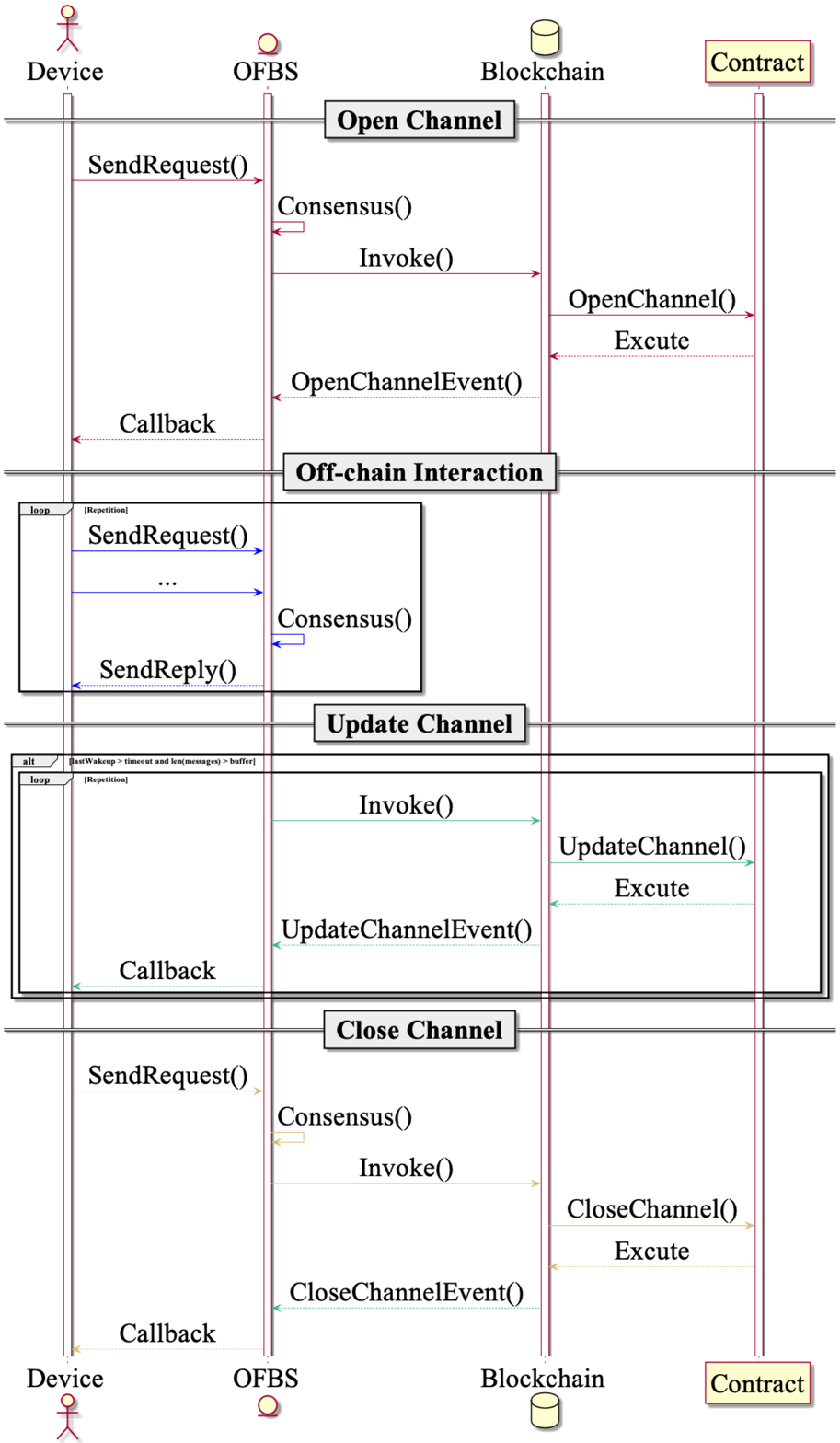}
  \caption{Workflow of OFBS}
  \label{workflow_ofbs}
  \end{figure}

Offline clients have little ability to connect to the Internet, which gives them no chance to interact with the blockchains. This section introduces the workflow of SMS-based OFBS to achieve interactions between offline clients and blockchains. OFBS is divided into on-chain and off-chain parts. For a better demonstration of OFBS, a piece of a sequence diagram is detailed in Fig.  \ref{workflow_ofbs}. \textit{Challenge 1: If a tree falls in the forest and no one is around to hear it, does it make a sound \cite{poon2016bitcoin}?}

\paragraph{Initialize} SMS service providers initialize infrastructures (like cell towers) as relay nodes for peer-to-peer networks. The initialization of infrastructure includes bandwidth, buffer size, wakeup time, time out, and other required information. Offline clients and relayers initialize private keys, public keys, account addresses, and certifications. The private key is used to sign transactions. The account address is used to identify clients and relayers uniquely. The digital certificate is used to access OFBS and blockchains. Moreover, SMS service providers are required to deploy channel contracts in advance on the consortium blockchains. 

The channel contracts are the intermediary that transfers state changes on the chain to off-chain to reduce the frequency of interaction with the blockchain. The functions of the contract include opening the channel, updating the channel, and closing the channel. The channel used here is different from the state channel \cite{poon2016bitcoin}. The state channel requires clients to be online, while the channel of OFBS can help offline clients access blockchains.

\paragraph{OpenChannel} Offline clients packages short messages according to the SMS protocol specification. The data packet includes the message content, destination number, encoding format, type, and other information. And the message content consists of the destination address, amount, signature, and timestamp. The data packet is sent to relayers.

In the traditional SMS process, after relayers receive the data packet, it forwards it to Short Message Center (SMC). Then SMC delivers it to the corresponding base station where the destination number is located. Finally, the base station forwards it to the connected device. 

In OFBS, the SMS-based data packet is forward to relay nodes. Since relay nodes are a peer-to-peer network, they need to execute consensus to maintain data integrity. Besides, the data packet is signed by the private key, and relay nodes can not modify the content. After relayers consensus, relayers invoke OPENCHANNEL(), and the data packet is written into the channel contract running on blockchain for consensus. The channel contract will escrow balances of offline clients for consequent transactions. After blockchain consensus, the channel contract will emit OpenChannel Event. Relay nodes monitor events and callback results to offline clients. See details in  Algorithm \ref{contract}.

\begin{algorithm}[!t]
  \caption{{Channel Contract}}
  \begin{algorithmic}[1] 
    \Function{OpenChannel}{client, relay, balance}
      \State channel[relay] = \{relay,balance, nonce=0, txs,...\}
      \State emit open channel event
    \EndFunction
    \State
    \Function{UpdateChannel}{relay, balance, tx, serial}
      \State channel[relay] = \{relay, balance, nonce, txs, ...\}
      \State channel.Update += 1
      \If{channel.Update == threshold}
        \State Aggregate(channel[relay])
        \State channel.Update == 0
        \State emit update channel event
      \EndIf 
    \EndFunction
    \State
    \Function{CloseChannel}{replayer, balance}
      \If {Aggregate(channel[relay])}
        \State refund and delete channel
        \State emit close channel event
      \EndIf
    \EndFunction
  \end{algorithmic}
  \label{contract}
\end{algorithm}

\paragraph{Off-chain Interaction} Since the channel is opened by OFBS, offline clients can directly interact with each other without trust. Offline clients initiate a SMS packet signed by the private key to relay nodes. Relay nodes execute consensus among nodes and select the leader. The leader broadcasts the message to workers for votes. Nodes that receive the message insert it into the database and push it to the queue. If the last wakeup time is bigger than timeout and the number of messages is bigger than the buffer size, relay nodes will proceed with the message to the blockchain to update the channel state. And the wakeup time is reset to the current moment. See details in Algorithm \ref{interaction}.

\begin{algorithm}[!t]
  \caption{{Off-Chain Interaction}}
  \begin{algorithmic}[1] 
    \State add modem(comport, baudrate, devid)
    \State add worker(modem, buffer, wakeup, timeout)
    \State add node(worker, mutex, requestPool, msgQueue)
    \State deploy channel contract
    \For {each node in nodes \textbf{in parallel}}
      \State send SMS request(uuid, mobile, message)
      \State execute consensus among base stations
      \State worker.EnqueueMessage(message, insertToDB=true)
      \If{wakeup $>$ timeout \& len(messages) $>$ buffer}
        \State modem.SendSMS(mobile, message)
        \State worker.EnqueueMessage(message, false)
        \State proceed message to on-chain consensus
        \State node.worker.wakeup = now
      \EndIf
    \EndFor
  \end{algorithmic}
  \label{interaction}
\end{algorithm}

Since offline clients can not directly query and modify the balance on the chain, it is hard for them to check the validity of transactions. Therefore, relay nodes need to ensure the transfer account has sufficient balance to support the transfer amount, check the validity of nonce to prevent replay attacks, and validate the signature of transactions.

\paragraph{Update Channel} OFBS sets the buffer pool and wakeup timeout to reduce the cost of frequent interaction with blockchains. If the update condition is met, messages on the queue are packed by relay nodes and written into the blockchain. We provide an on-chain aggregation method, while the off-chain aggregation method refers to Section \ref{ccbs_workflow}. Since we exploit consortium blockchains that provide high throughput and storage capabilities, reducing on-chain consumption costs is no longer our goal. Relay nodes invoke UPDATECHANNEL(), update amounts of offline clients, and emit UpdateChannel Event. Relay nodes monitor events and callback results to offline clients. See details in Algorithm \ref{contract}.

We support there are $\mathcal N$ relay nodes and $\mathcal F$ malicious nodes. The aggregation threshold is $\mathcal T$. We use $\mathcal X$ as a random variable to identify the number of malicious nodes in OFBS. Therefore, the probability of a faulty system is as follows.

\begin{equation}
	P[\mathcal X \geq \mathcal F] = \sum_{\mathcal X}^{\mathcal F} \frac{C_{\mathcal N - \mathcal F}^{\mathcal T - \mathcal X} C_{\mathcal F}^{\mathcal X}}{C_{\mathcal N}^{\mathcal T}}
\end{equation}

\paragraph{Close Channel} When offline clients decide to quit quick interaction with other clients, they can choose to close the channel. Offline clients initiate a SMS request signed by private keys to relay nodes. Relay nodes execute consensus and proceed messages to on-chain consensus. Relay nodes invoke CLOSECHANNEL()  to submit the latest off-chain states and aggregate states. Then they refund balances, delete channels, and emit CloseChannel Event. After monitoring events, they callback results to offline clients.

\section{Cross-chain Blockchain Service Design}
\label{ccbs_workflow}

Section \ref{ofbs-section} elaborates how offline clients interact with the blockchain. Next, we introduce CCBS to validate the authenticity of offline cross-chain transactions and propose a two-phase consensus to protect atomicity and integrity. \textit{Challenge 2: How do offline clients interact with Hyperledger Fabric and Xuperchain at the same time?}

\subsection{Overview}

\paragraph{CCBS} CCBS is a blockchain middleware for cross-chain transactions. It is based on OFBS to help offline clients forward and callback cross-chain transactions. CCBS interacts with the blockchain is to monitor proxy contracts deployed on the blockchain, including SourceContract and DestContract. CCBS has three types of nodes. Relay nodes are responsible for off-chain consensus and aggregation. The leader node is selected from relay nodes to monitor, broadcast, and aggregate cross-chain transactions. Apart from the duties of relay nodes, the monitoring nodes are the regulator of blockchains. These nodes can access real-time transactions to audit.

\paragraph{Destination Contract (DestChain)} DestChain is the destination for cross-chain transactions. A cross-chain transaction can get involved in more than one DestChain. Since CCBS exploits contracts as interfaces, it can redirect multiple transactions to many blockchains in one cross-chain transaction. Besides, DestChain can be the SourceChain, which depends on the relative relationship between blockchains.

\paragraph{Proxy Contract (SourceContract and DestContract)} The proxy contract is an interface for CCBS. For DestChain, the proxy contract is DestContract. Since one cross-chain transaction can be forwarded to multiple blockchains, there may be more than one DestContract deployed in blockchains. For SourceChain, the proxy contract is SourceContract. SourceContract receives cross-chain transactions, pushes them to the pending queue, and deletes them after the callback.

\paragraph{Source Blockchain (SourceChain)} SourceChain is the source of cross-chain transactions. Offline clients first initiate a cross-chain request through OFBS to SourceContract on the SourceChain. CCBS monitors the SourceContract and proceeds with corresponding operations. CCBS also monitors the DestContract on DestChain and callbacks related results to SourceChain.

\subsection{Workflow}

\begin{figure}[!t]
  \centering
  \includegraphics[width=3.5in, height=4in]{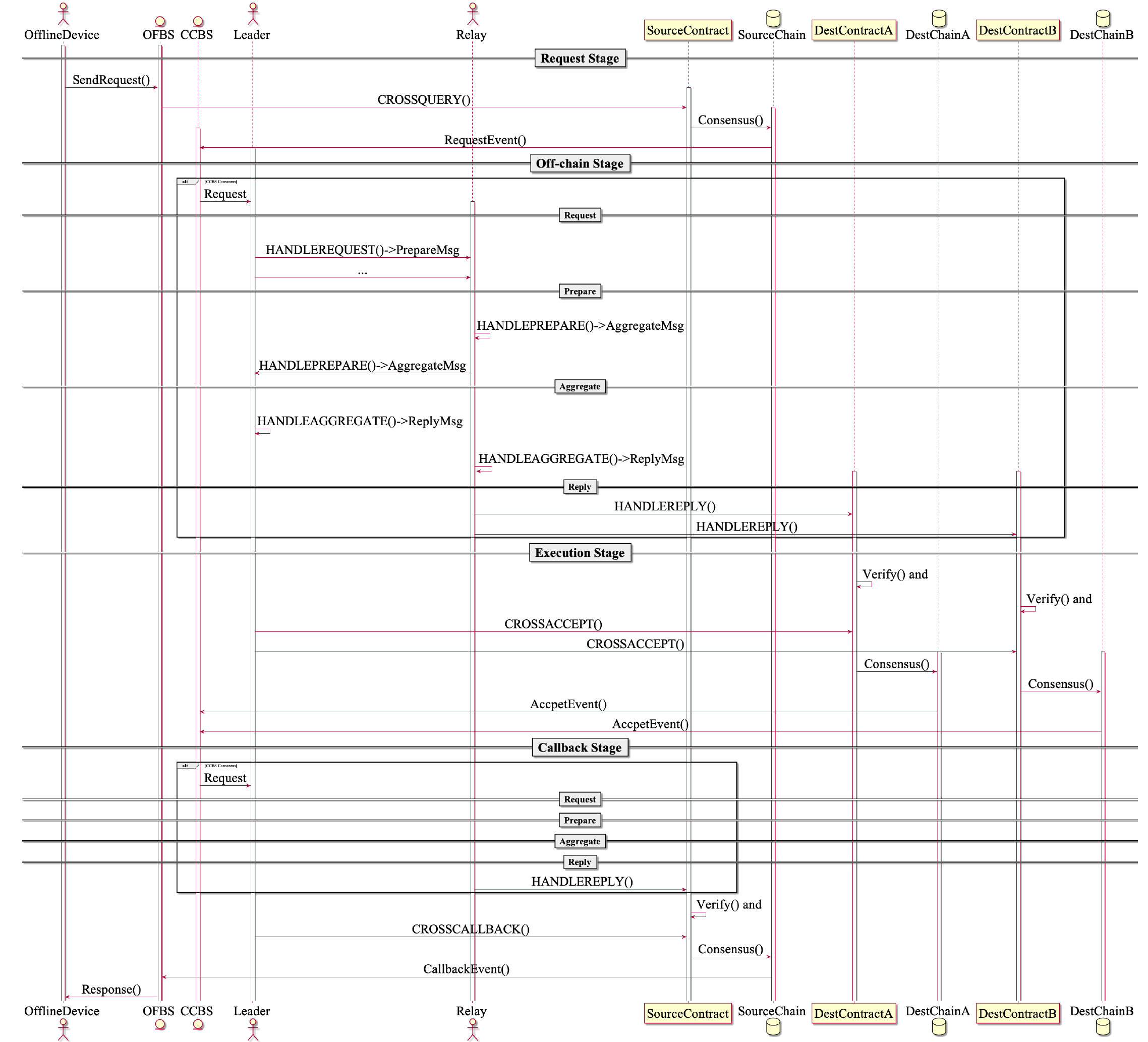}
  \caption{Workflow of CCBS}
  \label{workflow_ccbs}
  \end{figure}

The workflow of CCBS includes four main stages. For a better demonstration of CCBS, a piece of a sequence diagram is detailed in Fig. \ref{workflow_ccbs}. The main symbols used in the article are shown in Table \ref{symbol}.
 
\begin{table}[!t]
  \caption{MAIN SYMBOLS USED IN THIS PAPER}
  \begin{center}
  \resizebox{\linewidth}{!}{
  \Huge
  \begin{tabular}{|c|c|}
  \hline
  \textbf{Denotion} & \textbf{Description}\\
  \hline
      \hline
      $even_{k}$& requests events generated by virtual spaces\\
      \hline
      $sig$& Elliptic curve signatures\\
      \hline
      $apub_{th}$& Aggregated threshold public keys of all nodes\\
      \hline
      $subsig_{th}, subpub_{th}$ &  Aggregated threshold public keys of part nodes\\
      \hline 
      $S_b, D_b,C_p$ & SourceChain, DestChain, and Proxy Contract \\
      \hline
      $R_{i}, L_j$& Relay and Leader in CCBS\\
      \hline
      $g_1,g_2$ &Generators of multiplicative cyclic groups $G_1$ and $G_2$ \\
      \hline
      $e$ &The bilinear map: $G_1 \times G_2 \rightarrow G_3$ \\
      \hline
      $bsk_i,bpk_i, bsig_i$& The private key, public key and signature of nodes in Bilinear Aggregate Signature\\
      \hline
      $m,H$ & The message and hash function of messages\\
      \hline
  \end{tabular}}
  \label{symbol}
  \end{center}
\end{table}

\paragraph{Initialize} Offline clients initialize the public key, private key, account, certificate from blockchains or OFBS. Relay nodes $R_i$ of CCBS initialize the private key $bsk_i$, public key $bpk_i $ of Bilinear Aggregate Signature, and other required keys. Proxy Contract $C_p$ should be deployed in advance in multiple blockchains. And $R_i$ should broadcast $bpk_i$ to other $R_i$, aiming to  aggregate $apub_{th}$.  The aggregated public key $apub_{th}$ is written to $C_p$ in advance, which is used to verify the authenticity of results from $R_i$. 

\paragraph{Request Stage} An offline client initiates cross-chain transactions $request_k\mbox{(from, to, amount, data, ...)}$ through OFBS. OFBS determines if the type of $request_k$ is a cross-chain transaction, then forwards $request_k$ to CCBS. CCBS first invokes CROSSQUERY() to write $request_k$ into SourceContract of SourceChain $C_p^{S_b}$. The $request_k$ is added into pending queues and emits RequestEvent.

\paragraph{Off-chain Stage} Relay nodes $R_i$ elect a leader $L_j$ and $L_j$ listens to RequestEvent of $C_p^{S_b}$. Once $L_j$ listen for a non-empty message, $L_j$ extracts $request_k$ from RequestEvent. Then $L_j$ broadcast $request_k$ to other $R_i$ for the off-chain consensus.

\begin{algorithm}[!t]
  \caption{{Workflow}}
  \begin{algorithmic}[1] 
      \Require $request_{0...n}$ 
      \State $L_j \gets$ select from $R_i$ in consensus
      \State CROSSQUERY($request_{k}$) $\gets$ OFBS
      \State RequestEvent $\leftarrow $ Triggered by $C_p^{S_b}$
      \State $L_j$ listen to RequestEvent
      \For{each $request_{k} \in request_{0...n}$ \textbf{in parallel}}
          \State $L_j \rightarrow$ HANDLEREQUEST(payload)
          \For{$R_i$  \textbf{in parallel}}
              \State $R_i\rightarrow$  HANDLEPREPARE()
              \State $R_i\rightarrow$  HANDLEAGGRE()
               \State $R_i\rightarrow$  HANDLEREPLY()
          \EndFor
          \State $L_j \rightarrow$ CROSSACCEPT() in $C_p^{D_b}$
          \State $L_j \rightarrow$ CROSSCALLBACK() in $C_p^{S_b}$
      \EndFor
  \end{algorithmic}
  \label{negotiation}
\end{algorithm}

The off-chain consensus process is divided into four stages, including \textit{Request, Prepare, Aggregate, and Reply stage}. Algorithms \ref{oracle_consensus} shows the specific process. It should be noted that the following process can be adapted into multiple DestChain as shown in Fig. \ref{workflow_ccbs}. For convenience, we only show a single DestChain in the following process.

\begin{algorithm}[!t]
  \caption{{Off-chain Stage}$\quad\triangleright \textrm{Run on CCBS}$}
  \begin{algorithmic}[1] 
      \Function{HandleRequest}{payload}
          \State $request_k  \gets $ RequestEvent
          \For{each $request_{k} \in request$ \textbf{in parallel}}
              \State prepareMsg $\gets$  $request_k$
              \State Broadcast(Sign(prepareMsg, $sk_l$)) 
          \EndFor
      \EndFunction
      \State
      \Function{HandlePrepare}{payload}
          \For{$R_i \in$ Relay Nodes \textbf{in parallel}}
              \If{Verify(prepareMsg,$sig_{l}, pk_{l}$)} 
              	\State aggregateMsg $\gets$ Params($request_k$)
              	\State Broadcast(Sign(aggregateMsg, $bsk_i$)) to $R_i$              
              \EndIf
          \EndFor
      \EndFunction
      \State
      \Function{HandleAggregate}{payload}
          \For{$R_i \in$ Relay Nodes \textbf{in parallel}}
          	\State collect $bsig_i$ of aggregateMsg
          	\If{$e(g_1,bsig_i) == e(bpk_i, H)$}
          		\State log.append(aggregateMsg, $bsig_i$, $bpk_i$)
          	\EndIf
          	\State $ subsig_{th} \gets \prod_{i=1}^{w}bsig_i, subpub_{th} \gets \prod_{i=1}^{w}bpk_i$
          \EndFor
      \EndFunction
      \State
      \Function{HandleReply}{payload}
      	\State provide ($request_k, subsig_{th}, subpub_{th}$)
      	\State invoke CROSSACCEPT() of $C_p^{D_b}$
      \EndFunction
  \end{algorithmic}
  \label{oracle_consensus}
\end{algorithm}

\begin{itemize}
	\item \textit{Request:} $L_j$ executes HANDLEREQUEST(), and broadcasts prepareMsg to $R_i$. $L_j$ signs prepareMsg by the private key so that $R_i$ can verify the authenticity by the public key of $L_j$.

	\item \textit{Prepare:} $R_i$ executes HANDLEPREPARE() to verify prepareMsg from $L_k$. If verified, $R_i$ signs aggregateMsg by $bsk_i$ and broadcast aggregateMsg to other $R_i$, as shown in \eqref{prepare-1}-\eqref{prepare-3}.
	
  \begin{equation}
    g_i^{bsk_i} \rightarrow bpk_i
    \label{prepare-1}
  \end{equation}    
  \begin{equation}
    H(m) \rightarrow H, H^{bsk_i} \rightarrow bsig_i \in G_2
    \label{prepare-2}
  \end{equation}  
  \begin{equation}
    e(g_1,bsig_i) = e(g_1^{bsk_i},H)=e(bpk_i, H)
    \label{prepare-3}
  \end{equation}  

	\item \textit{Aggregate:} $R_i$ collects $bsig_i$ of aggregateMsg and verify the authenticity through $bpk_i$. If $R_i$ receives enough aggregateMsgs of same results, $R_i$ executes HANDLEAGGREGATE() to aggregate $bsig_i$ and get the aggregated signature $subsig_{th}$ and public key $subpub_{th}$, as shown in \eqref{agg-1}-(\ref{agg-3}).
	
  \begin{equation}
    H_i=H(m_i),i=1,2,…,w
    \label{agg-1}
  \end{equation}  
  \begin{equation}
    subsig_{th} = \prod_{i=1}^{w}bsig_i
    \label{agg-2}
  \end{equation}  
  \begin{equation}
    e(g_1,subsig_{th})=\prod_{i=1}^{w}e(g_1^{bsk_i},H_i)
    \label{agg-3}
  \end{equation}  

  \item \textit{Reply:} $R_i$ executes HANDLEREPLY() to invoke CROSSACCEPT() of DestContract of DestChain $C_p^{D_b}$. It provides $request_k$, $subsig_{th}$  and $subpub_{th}$ to $C_p^{D_b}$.
  
\end{itemize}

\begin{algorithm}[!t]
  \caption{{Proxy Contract}$\quad\triangleright \textrm{Run on blockchains}$}
  \begin{algorithmic}[1] 
      \Function{CROSSQUERY}{payload}
      \State Accept $request_{k}$ from OFBS
      \State $apub_{th} \gets$ constructor()
      \State $pending[req] \gets (request_{k}, apub_{th})$
      \State Emit RequestEvent
      \EndFunction
      \State
      \Function{CROSSACCEPT}{payload}
      \State Accept $(request_{k}, subsig,subpub, mask)$ from $R_i$
      \State $apub_{th} \gets$ constructor()
      \State $\mbox{multiSig} \gets \mbox{bls.Multisig}(subsig_{th}, subpub_{th}, mask)	$
      \If{$\mbox{multiSig.Verify}(apub_{th}, request_k)$}
          \State Accept $(request_{k}$ and modify state
      \EndIf
      \State Emit \textit{AccpetEvent}
      \EndFunction
      \State
      \Function{CROSSCALLBACK}{payload}
      \State Accept $(result_{k}, subsig,subpub, mask)$ from $R_i$
      \State $(request_{k}, apub_{th}) \gets pendingReq[req]$
      \State $\mbox{multiSig} \gets \mbox{bls.Multisig}(subsig_{th}, subpub_{th}, mask)	$      
      \If{$\mbox{multiSig.Verify}(apub_{th}, result_k)$}
          \State Delete $request_{k}$ from $pending[req]$
      \EndIf
      \State Emit \textit{CallbackEvent}
      \EndFunction
  \end{algorithmic}
  \label{proxy_contract}
\end{algorithm}

\paragraph{Execution Stage} $C_p^{D_b}$ exploits public keys of all relay nodes $apub_{th}$ to verifiy $(request_k, subsig_{th}, subpub_{th})$, as shown in \eqref{agg-4}-(\ref{agg-5}). Mask represents the $R_i$ who really signed. If verified, $C_p^{D_b}$ execute operations and emits AcceptEvent.

\begin{equation}
	\mbox{multiSig} \gets \mbox{bls.Multisig}(subsig_{th}, subpub_{th}, mask)	
	    \label{agg-4}
\end{equation}
\begin{equation}
	\mbox{multiSig.Verify}(apub_{th}, request_k)
	    \label{agg-5}
\end{equation}

\paragraph{Callback Stage} $L_j$ listens to AcceptEvent of $C_p^{D_b}$ and extracts the callback result $result_k$ from $event_k$. Then $L_j$ broadcasts $result_k$ to $R_i$ and executes Off-chain Stage. $R_i$ executes HANDLEREPLY() to invoke CROSSCALLBACK() of $C_p^{S_b}$ and provide $result_k$, $subsig_{th}$  and $subpub_{th}$ to $C_p^{S_b}$. $C_p^{S_b}$ verifies the authenticity of $result_k$, delete $request_k$ from pending queue and emit CallbackEvent. OFBS listens to $event_k$ and response $result_k$ to offline clients.

The above process is as shown in Algorithm \ref{negotiation}, \ref{proxy_contract} and \ref{oracle_consensus}.

\section{Computing Blockchain Service Design (CPBS)}
\label{cpbs_workflow}

In this section, we introduce CPBS to help offline clients implement complex queries and computation on the blockchains. \textit{Challenge 3: How can offline clients analyze account activity using on-chain data from the past few months?}

\subsection{Overview}

\paragraph{CPBS} CPBS is a blockchain middleware implementing complex queries and computations for offline clients. CCBS is composed of multiple relay nodes and constructs a peer-to-peer network. It is based on OFBS to help offline clients forward and callback results. It is also based on CCBS to help offline clients forward and callback results from multiple blockchains. Therefore, when offline clients only specify the data source involving one chain, it only needs to combine OFBS and CPBS. Otherwise, it needs to combine OFBS, CCBS, and CPBS to obtain results.

\textit{Single On-Chain Source} Offline clients initiate tasks to OFBS, and OFBS forwards them to CPBS. After request, execute, aggregate, reply stages, relay nodes of OFBS callback results to OFBS. Finally, OFBS callbacks results to offline clients. See details in Section \ref{ocbs-workflow}.

\textit{Multiple On-Chain Sources} The process of multiple on-chain sources is similar to the single on-chain source. The difference is that when data sources of tasks get involved in multiple blockchains, relay nodes exploit CCBS to extract data from multiple blockchains. 

\paragraph{CompChain \& CompContract} CompContract $C_c$ is an interface to receive tasks from offline clients, while CompChain $C_b$ is used for executing CompContract. Relay nodes of CPBS monitor events of CompContract to see if there are unresolved tasks.

\paragraph{DestChain} DestChain is the data source specified by offline clients. Offline clients can specify multiple blockchains as data sources. Therefore, there may be more than one DestChain.

\paragraph{Offline device} Offline clients issue tasks through OFBS. For example, they can command CPBS to analyze specified account activity using on-chain data from the past few months.

\subsection{Workflow}
\label{ocbs-workflow}

\begin{figure}[!t]
  \centering
  \includegraphics[width=3.5in, height=4in]{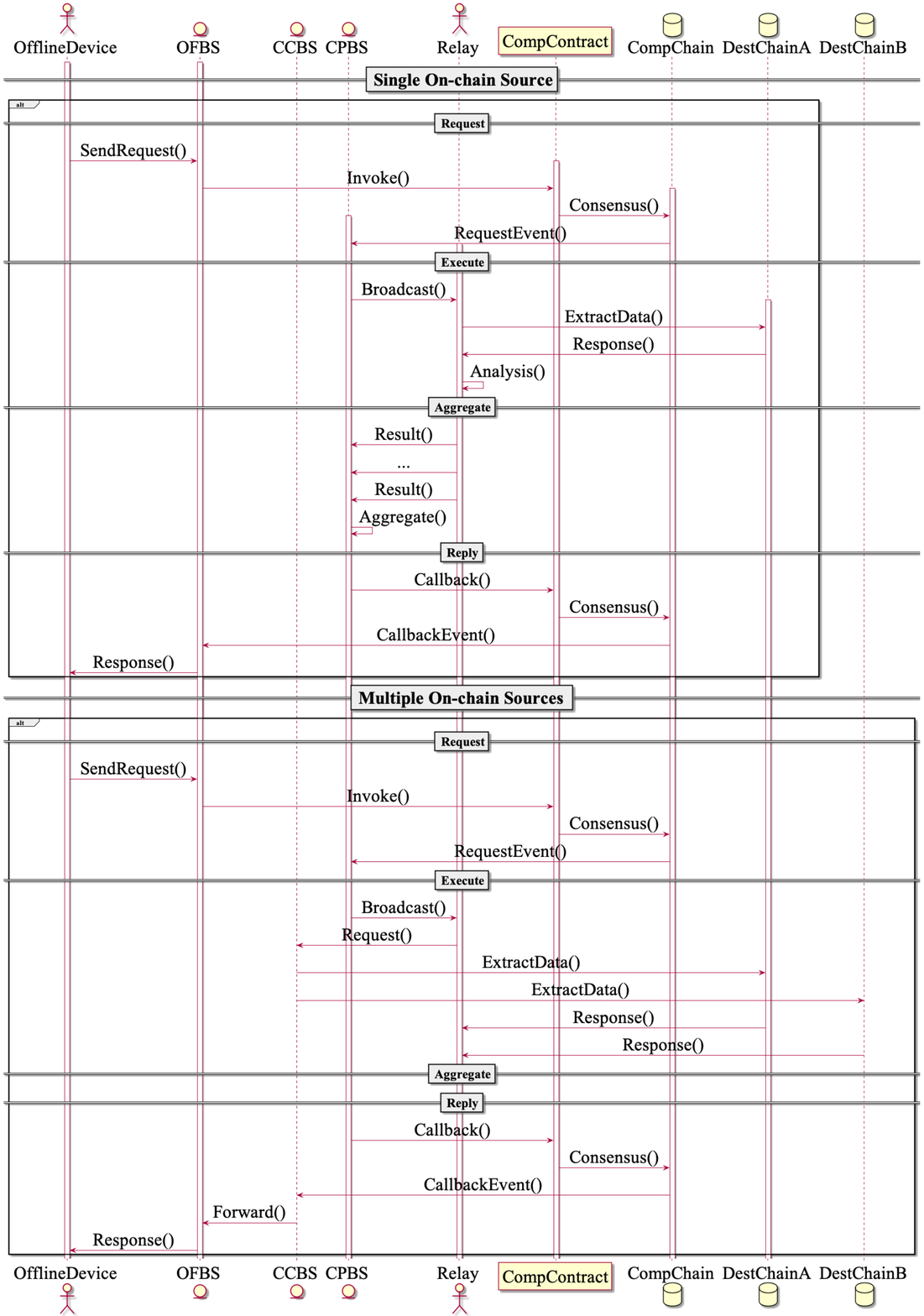}
  \caption{Workflow of CPBS}
  \label{workflow_cpbs}
  \end{figure}

The workflow of CPBS includes four stages. For a better demonstration of CPBS, a piece of a sequence diagram is detailed in Fig. \ref{workflow_cpbs}. 

\paragraph{Request} Offline clients initiate $task_k$ to OFBS. Since offline clients interact with OFBS through SMS, we can predefine some classic tasks so that offline clients can enter numbers to select tasks. For example, offline clients can enter number 1 to analyze account activities. Then OFBS forwards $task_k$ to CompContract $C_c$ and CompChain $C_b$ emits RequestEvent.

\paragraph{Execute} CPBS listens to RequestEvent and broadcast $task_k$ to relay nodes $R_i$. $R_i$ gets the data from the specified data source after it parses out $task_k$. If the specified data source gets involved in multiple blockchains, $R_i$ exploits CCBS to obtain data. After $R_i$ obtains data from data sources, it executes $task_k$ and output $result_k$. $R_i$ signs $result_k$ with $bsk_i$ and broadcasts $result_k$, $bsig_i$ and $bpk_i$ to other $R_i$.

\paragraph{Aggregate} $R_i$ collects $result_k$, $bsig_i$ and $bpk_i$. If $R_i$ receives enough results, $R_i$ aggregate the results and obtains $subpub_{th}$ and $subsig_{th}$, as shown in \eqref{agg-1}-(\ref{agg-3}).

\paragraph{Reply} $R_i$ writes $result_k$, $subpub_{th}$ and $subsig_{th}$ into $C_c^{C_b}$. $C_c^{C_b}$ verifies results, as shown in \eqref{agg-4}-(\ref{agg-5}). If verified, $C_c^{C_b}$ emits CallbackEvent. OFBS listens to CallbackEvent and response $result_k$ to offline clients.

\section{Experimental Evaluations}

This section evaluates the performance of the BcMON. First, we conduct a security analysis. Second, we evaluate the query performance of blockchain clients under poor network connections. Third, we evaluate the overhead of OFBS, CCBS, and OCBS.

\subsection{Security Analysis}

\paragraph{Illegal Transaction} Since offline clients have little chance to connect to the Internet. It is hard to determine whether offline clients have enough balance to pay the transfer amount. And it is hard to verify the signatures of transactions and whether the transaction is not a replay transaction. BcMON constructs three types of blockchain middleware to connect clients and blockchains. And the blockchain middlewares are required to execute consensus when submitting results.

\paragraph{Modify transactions of offline clients} Relay nodes of OFBS may do evil and modify transactions of offline clients. Since transactions are signed by offline clients, the tampered transaction will not pass verifications on the blockchain.

\paragraph{Ignore transactions of offline clients} Relay nodes of OFBS may intentionally ignore transactions of some offline clients. In general, relay nodes are represented by cell towers, which are no incentive to ignore transactions of offline clients. However, offline clients can replace SMS providers to change relay nodes if this happens. Or they can quit the network when there is an occasional internet connection.

\paragraph{Update old states of OFBS} Since OFBS exploits the channel to reduce interactions with blockchains, relay nodes of OFBS may update old states to the channel contracts. To prevent this issue, OFBS uses the on-chain aggregation for consensus among relay nodes. Therefore, the final state of the channel is the latest.

\paragraph{Atomicity and Consistency of CCBS} Offline clients are hard to determine whether the transaction succeeds or fails on blockchains. They also do not ensure consistency on blockchains. Therefore, CCBS constructs a two-phase consensus to keep atomicity and consistency.

\paragraph{Correctness and completeness of CPBS} There are malicious relay nodes of CPBS that return incorrect results. CPBS is different from Oracle \cite{mammadzada2019blockchain}, which may yield different results. Since the data sources come from blockchain, they are deterministic data. Therefore,  Correct results can be aggregated as long as a sufficient number of honest nodes return the same correct results, verifying the correctness and completeness.

\subsection{Query Performance Under Poor Connection}
\label{exp_query}

We exploited Xuperchain V3.10\footnote{https://github.com/xuperchain/xuperchain.git} to construct a local blockchain network. We queried the account, block, and transaction through the blockchain client named xclient. Each xclient is a grpc connect. Xuperchain is deployed on macOS Catalina 10.15.4, CPU 2.3 GHz Intel Core i5 with two cores, 16 GB 2133 MHz LPDDR3, 304.2 Mbit/s, and Go1.17.1. We initiated query requests concurrently through GoRoutine to calculate the throughput and total time. Besides, we also adjusted network connection (device, latency $\tau$, bandwidth $\beta$, and packet loss $\iota$) through comcast\footnote{https://github.com/tylertreat/comcast}. Based on the above settings, we simulated four network states (DEFAULT, WIFI, EDGE, and GPRS)  to evaluate the impact of the network performance of blockchain. The parameters are shown in Table \ref{parameters}.

\begin{table}[!t]
  \centering  
	\caption{Simulation Parameters}  
	\label{parameters}  
  \begin{tabular}{c|c}
  \hline
  Parameter & Value \\
  \hline
  \hline
  DEFAULT & Default congifuration\\
  \hline
  WIFI & $\tau$=40, $\beta$=30000, $\iota$=0.2 \\
  \hline
  EDGE & $\tau$=300, $\beta$=250, $\iota$=1.5 \\
  \hline
  GPRS & $\tau$=500, $\beta$=50, $\iota$=2 \\
  \hline
  Concurrent clients & DEFAULT=100, WIFI=20, EDGE=10, GPRS=5 \\
  \hline
  \multirow{4}{*}{MaxCount} & DEFAULT={[}1000,10000,1000{]}   \\
                          & WIFI={[}40,200,40{]},           \\
                          & EDGE={[}40,120,20{]}              \\
                          & GPRS={[}9,45,9{]}               \\
  \hline
  \multirow{4}{*}{Throughput} & DEFAULT={[}100000,1000000,100000{]}   \\
  & WIFI={[}800,4000,800{]},           \\
  & EDGE={[}400,1200,200{]}              \\
  & GPRS={[}45,225,45{]}               \\
  \hline
  Target protocol & tcp,udp,icmp \\
  \hline
  Device & eth0 \\
  \hline 
  \hline
  \end{tabular}
\end{table}

\begin{figure}[!t]
  \centering
  \subfigure[DEFAULT]{\includegraphics[width=1.5in]{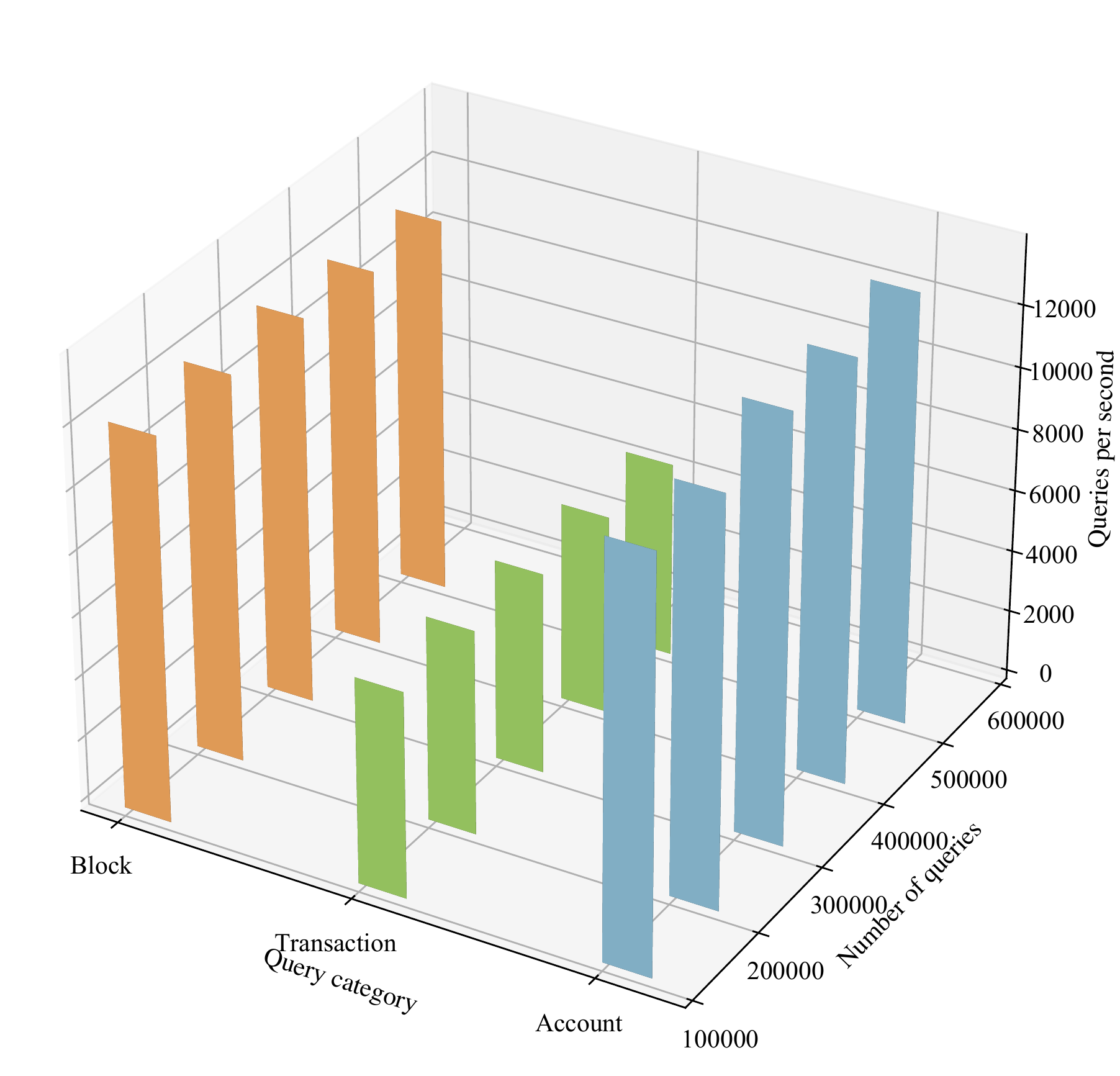}}
  \subfigure[WIFI]{\includegraphics[width=1.5in]{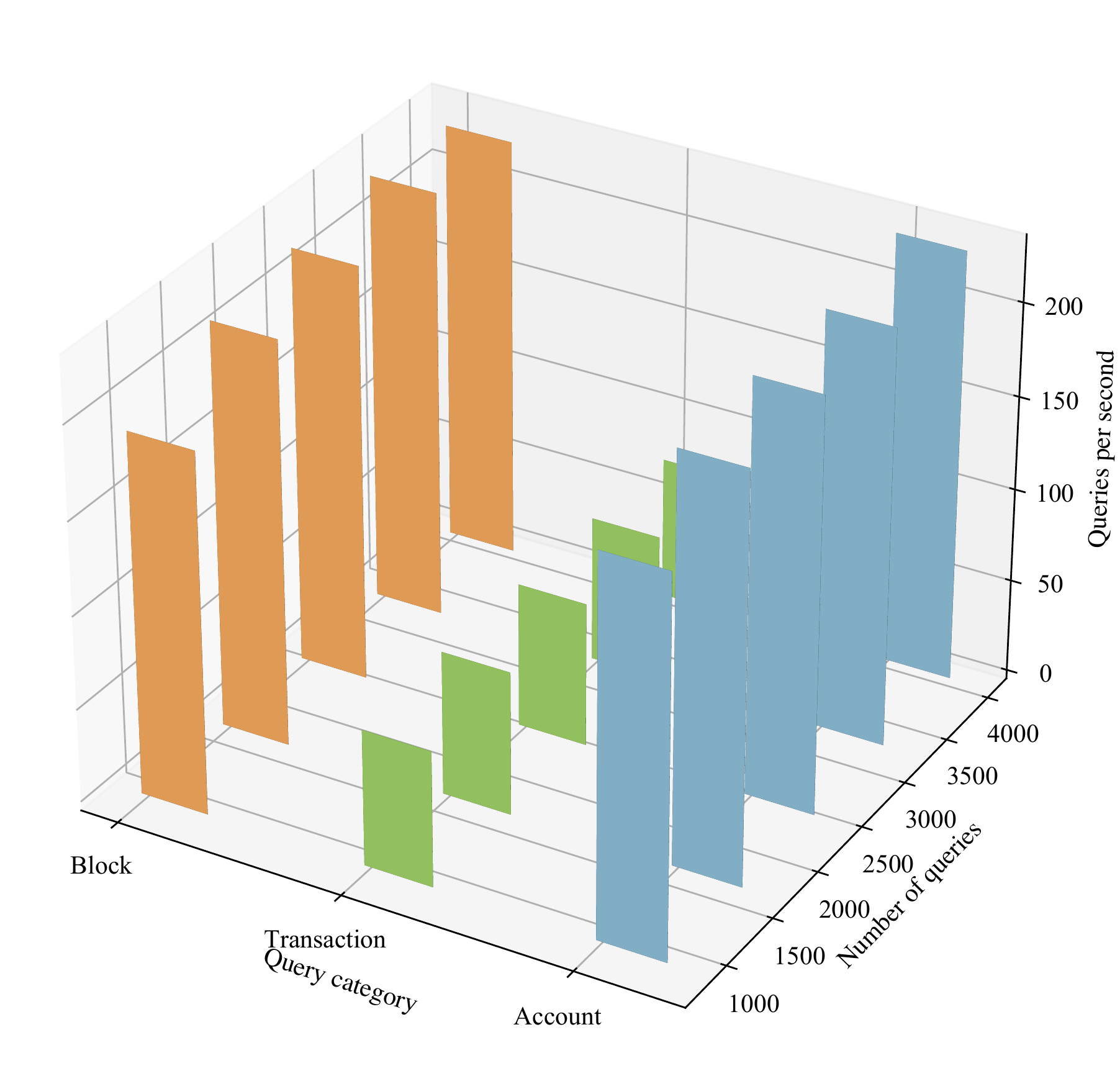}}
  \subfigure[EDGE]{\includegraphics[width=1.5in]{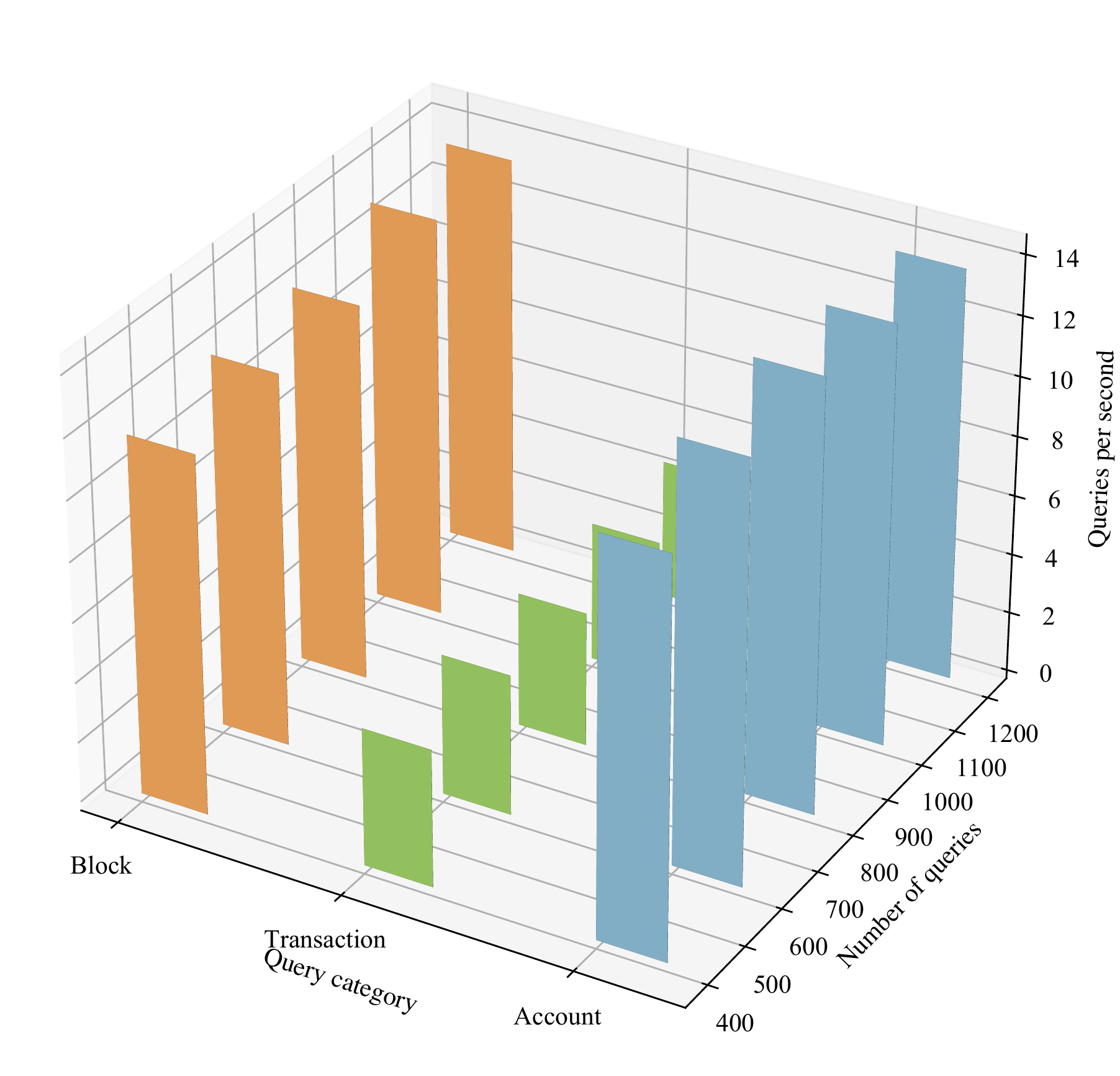}}
  \subfigure[GPRS]{\includegraphics[width=1.5in]{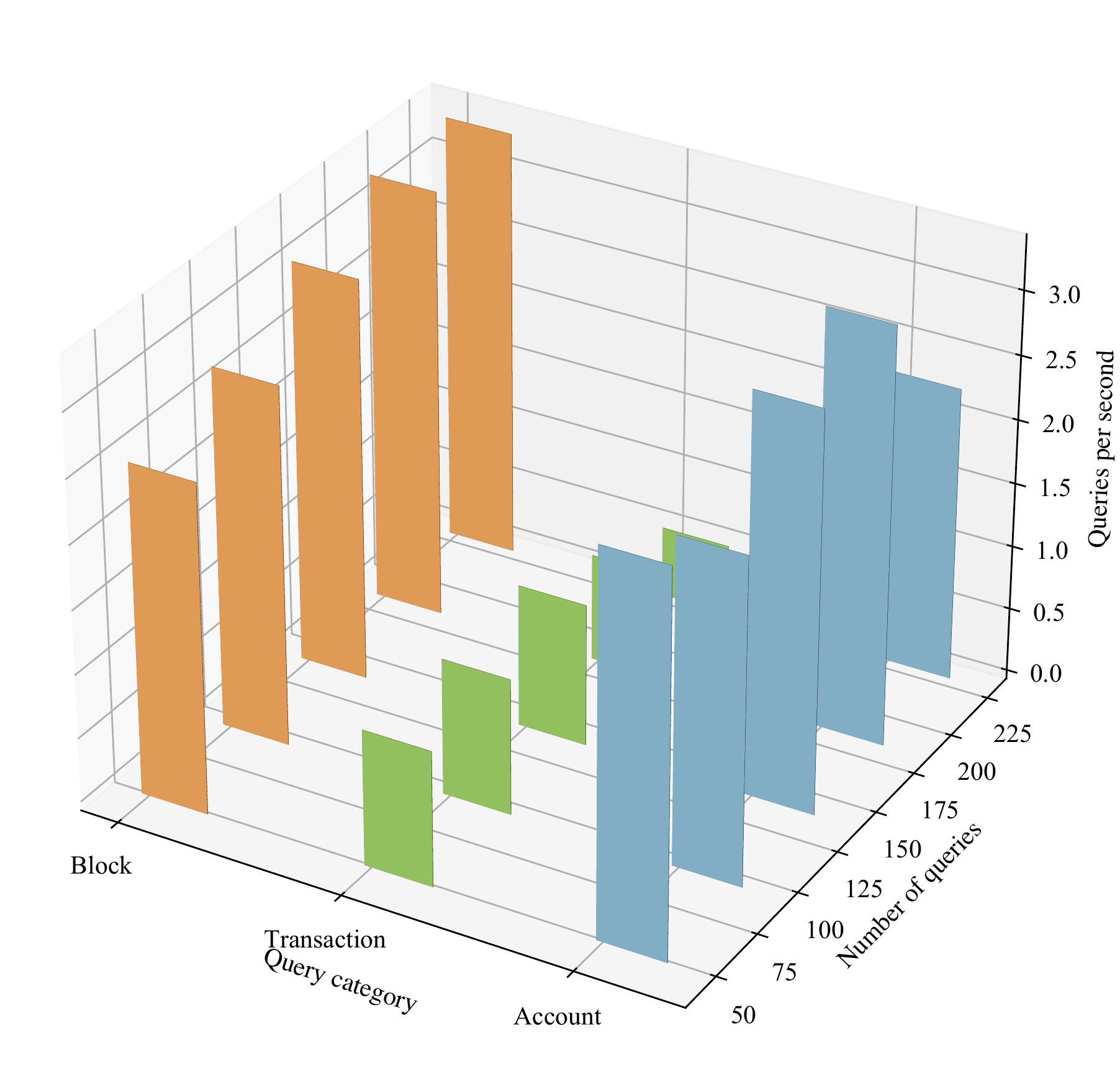}}
  \caption{Throughput}
  \label{xclient_network_throughput}
\end{figure}

\begin{figure}[!t]
  \centering
  \subfigure[DEFAULT]{\includegraphics[width=1.5in]{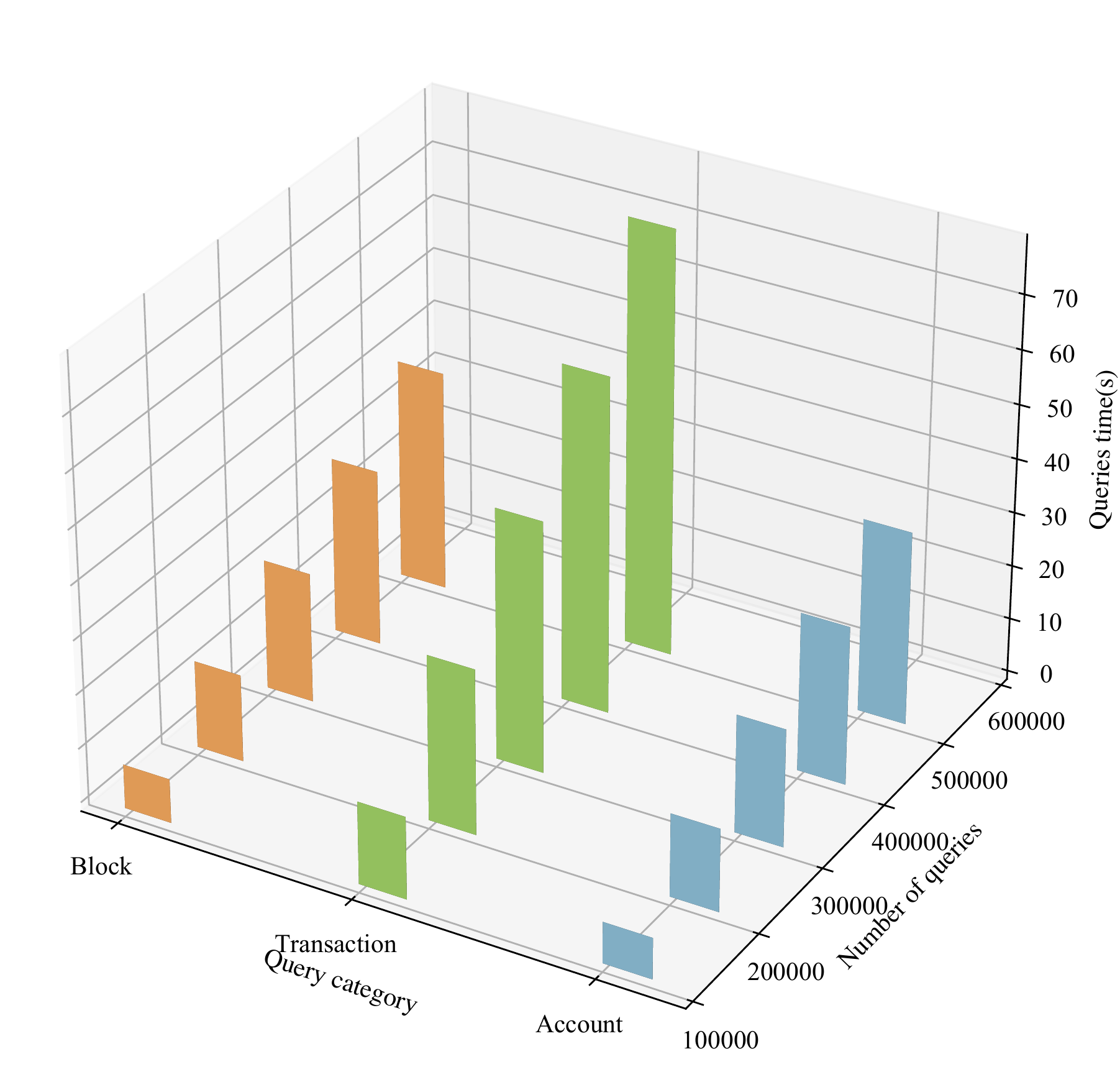}}
  \subfigure[WIFI]{\includegraphics[width=1.5in]{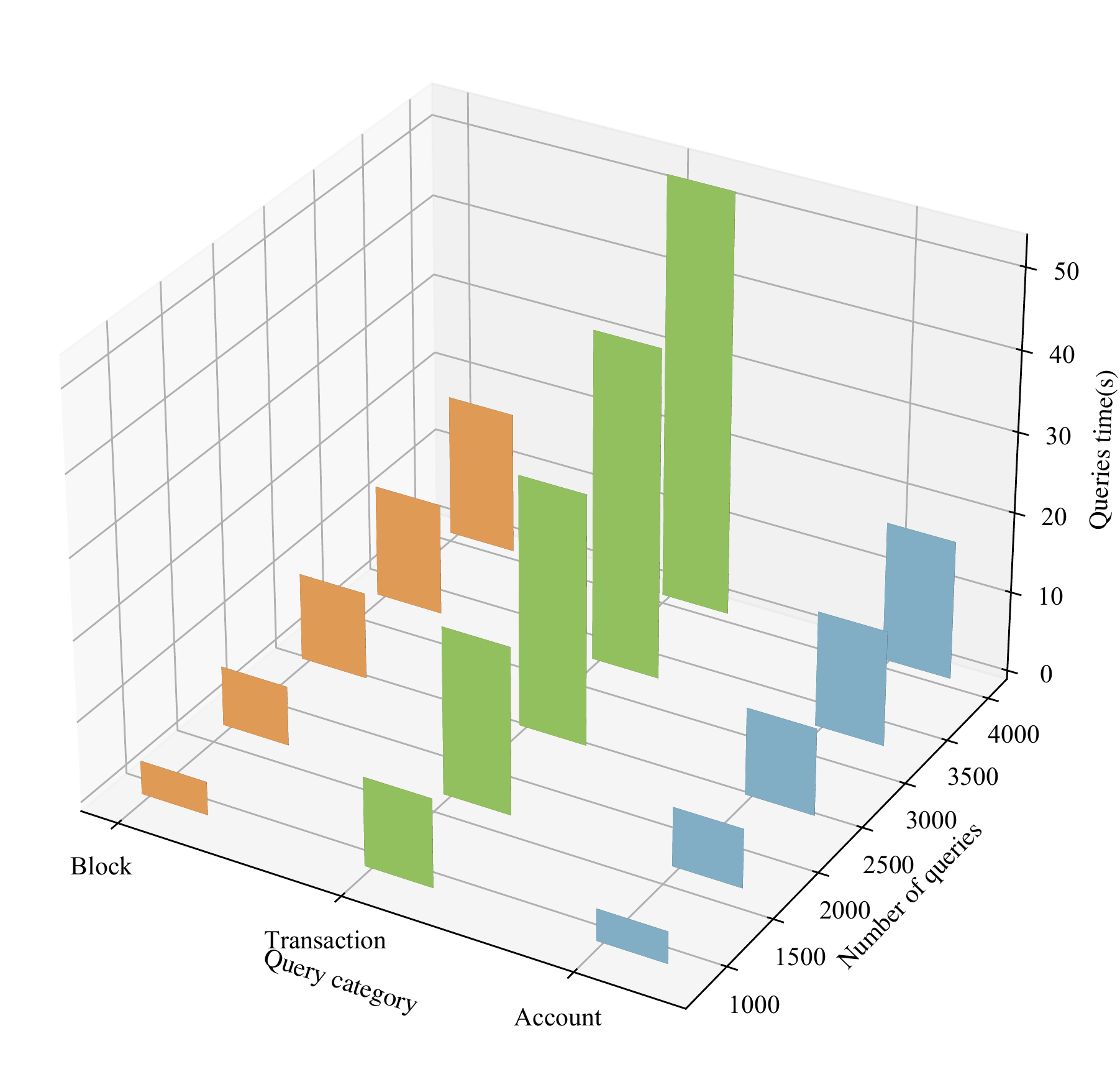}}
  \subfigure[EDGE]{\includegraphics[width=1.5in]{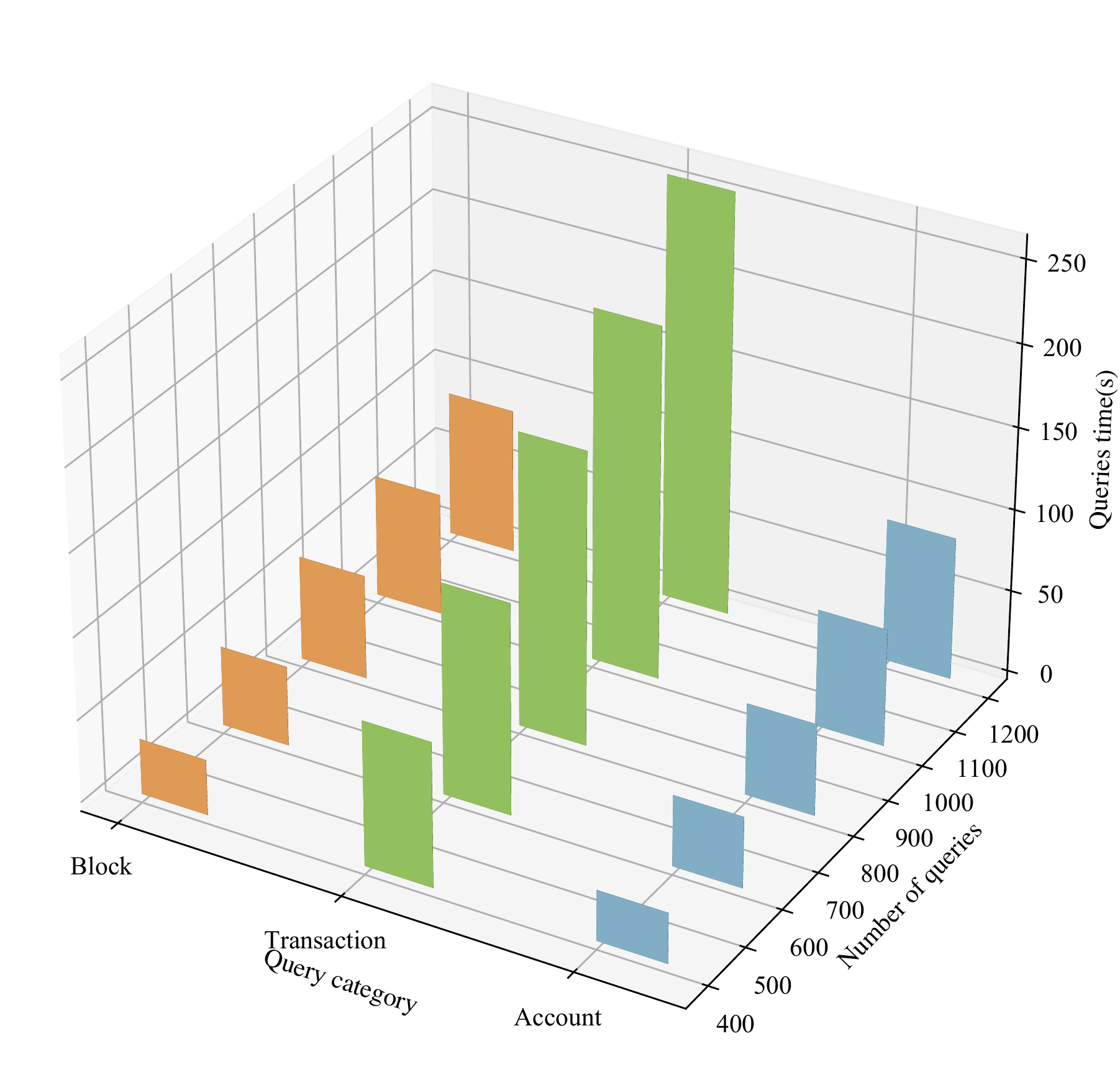}}
  \subfigure[GPRS]{\includegraphics[width=1.5in]{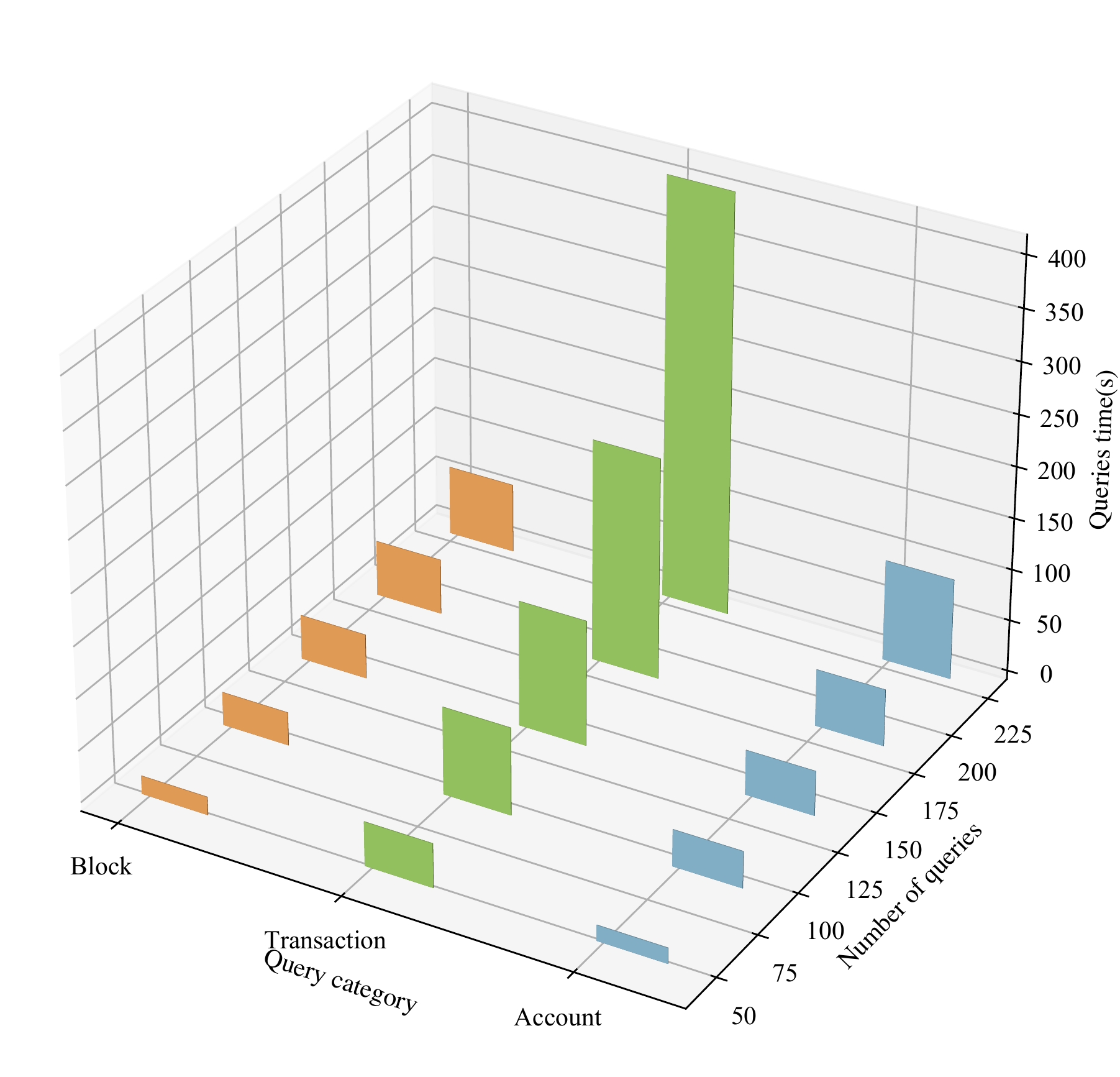}}
  \caption{Query Time}
  \label{xclient_network_time}
\end{figure}

The asynchronous query requests exceed the QPS (Queries Per Second) by two orders of magnitude, avoiding the impact of boundary situations. Fig. \ref{xclient_network_throughput} and \ref{xclient_network_time} are the throughput and query time of different network performances. The QPS used by DEFAULT (Fig. \ref{xclient_network_throughput} and Fig. \ref{xclient_network_time} (a) is unbearable for WIFI, EDGE and GPRS (Fig. \ref{xclient_network_throughput} and Fig. \ref{xclient_network_time} (b) (c) (d)). And the network resources spent in querying blocks and accounts are about 50\% of the query transactions. In summary, blockchain technology cannot be connected to the vast majority of people worldwide.

\subsection{Overhead of OFBS}
\label{exp_ofbs}

The configuration related to Xuperchain is the same as Section \ref{exp_query}. We exploited contract-sdk-go to implement the channel contract. The database for SMS is sqlite3. OFBS was deployed on macOS Catalina 10.15.4, CPU 2.3 GHz Intel Core i5 with two cores, 16 GB 2133 MHz LPDDR3, and 304.2 Mbit/s. We exploited PBFT \cite{castro1999practical} for the off-chain consensus of OFBS. The on-chain consensus of Xuperchain is the default configuration of the SINGLE consensus. Since OFBS is the blockchain middleware, it is irrelevant to the on-chain consensus. Moreover, Xuperchain supports the pluggable consensus. Therefore, we choose the simplest on-chain consensus during the experiment. 

We divided the experiments into the off-chain and the on/off-chain parts to evaluate the performance of OFBS. The off-chain part is relevant with OFBS. It needs to evaluate the off-chain service time when the number of relay nodes and concurrent clients increases. Since the scenarios OFBS applied are limited, we only exploited a small number of relay nodes [4, 9, 1] to participate in the experiment. The on/off-chain part includes the off-chain part and the on-chain part. It needs to evaluate the total service time when the number of relay nodes, concurrent clients, and the contract method is different.

\begin{figure}[!t]
  \centering
  \subfigure[Off-Chain]{\includegraphics[width=1in]{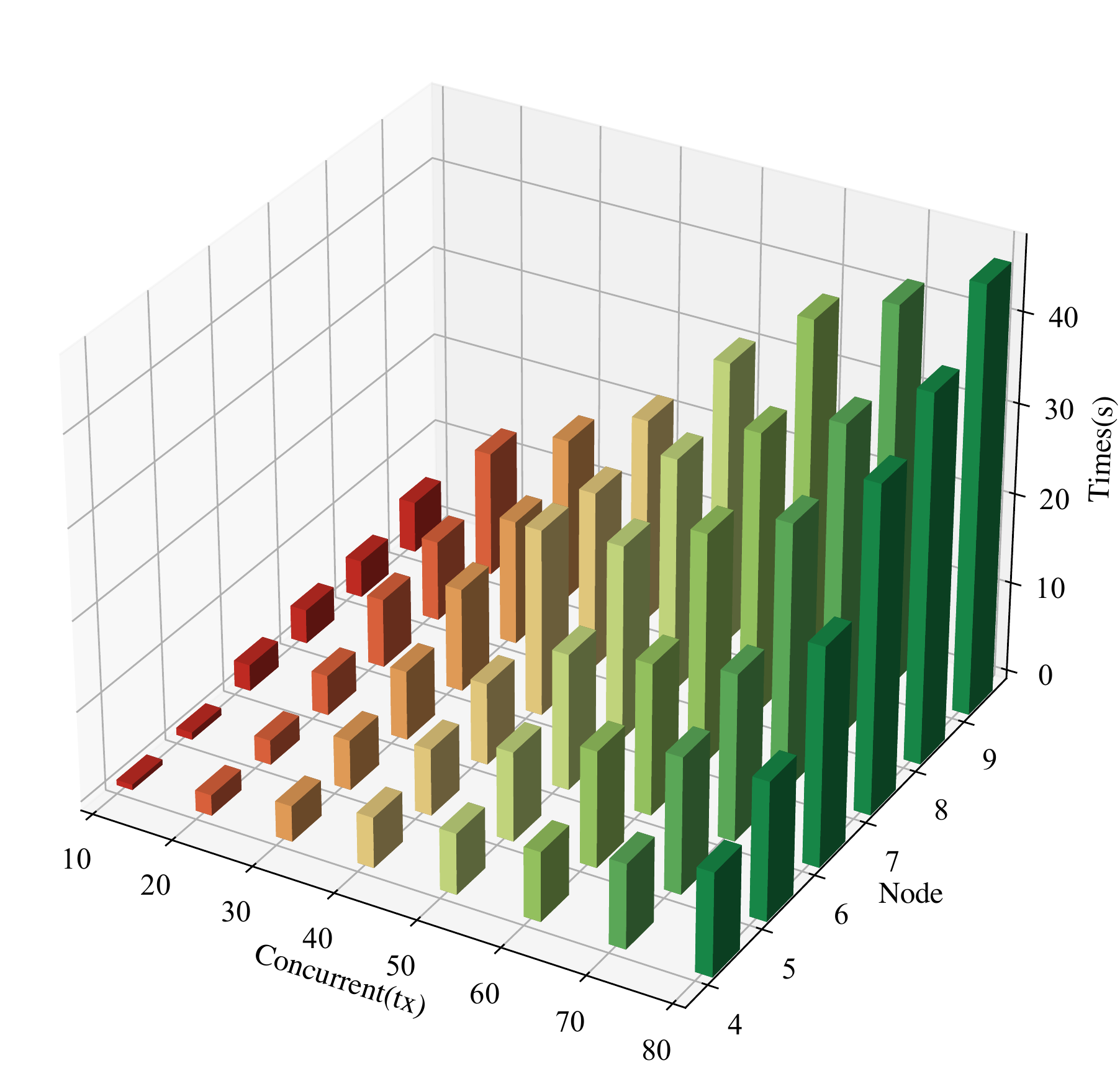}}
  \subfigure[OpenChannel]{\includegraphics[width=1in]{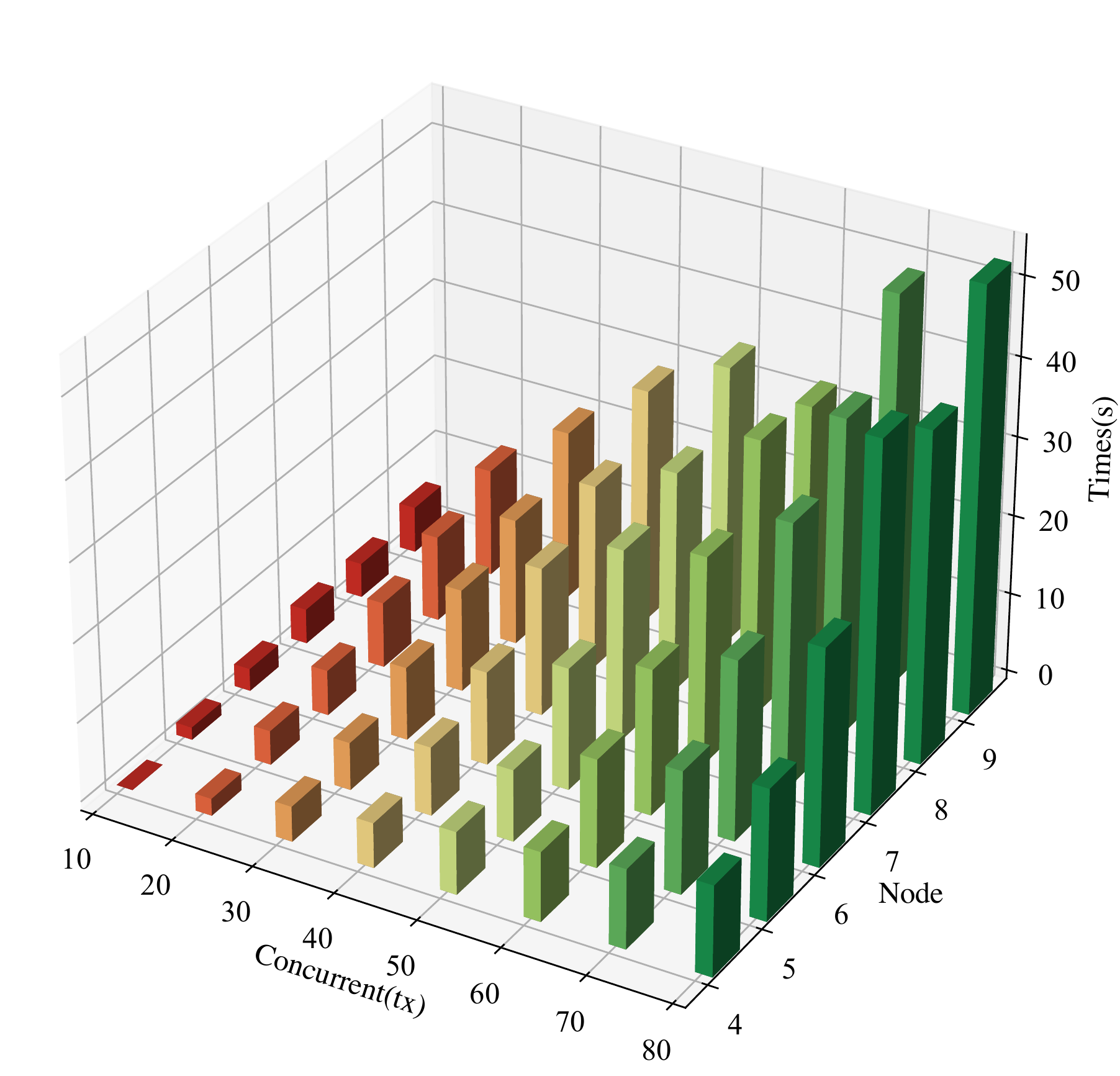}}
  \subfigure[UpdateChannel]{\includegraphics[width=1in]{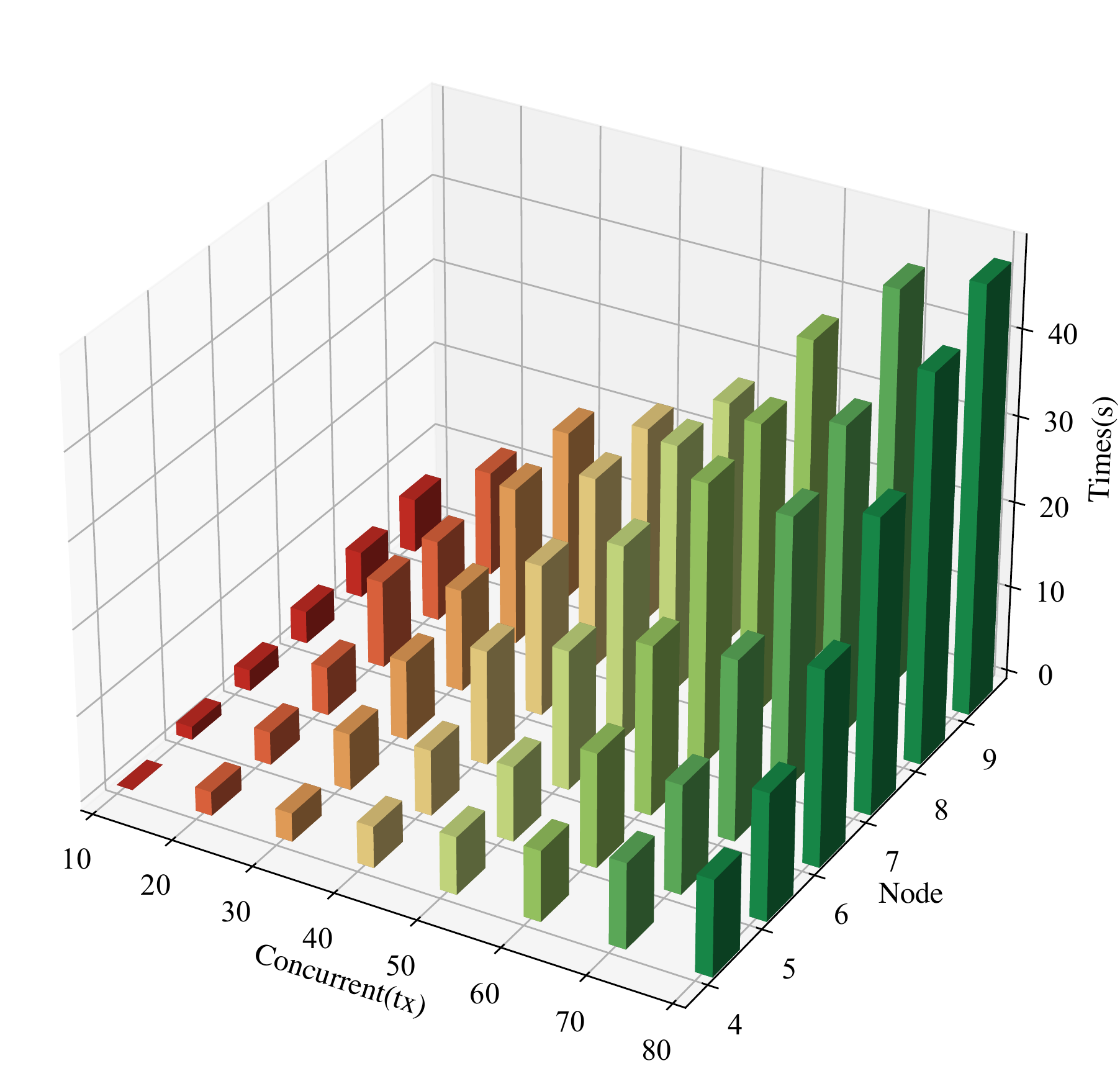}}
  \caption{Service ime}
  \label{chain-time}
  \end{figure}

Fig. \ref{chain-time} (a) shows the off-chain service time when the number of relay nodes is [4, 9, 1] and the number of concurrent clients is [10, 80, 10]. As can be seen from the figure, as the number of relay nodes increases, the average service time per transaction increases. This reason is that the off-chain SMS-based relay nodes need to access the database frequently. Fig. \ref{chain-time} (b) and (c) show the total service time when the number of relay nodes is [4, 9, 1], the number of concurrent clients is [10, 80, 10], and the contract methods are OpenChannel and UpdateChannel. The concurrent clients initiate OpenChannel and UpdateChannel requests to modify the state of the blockchain. And relay nodes of OFBS forward requests to the blockchain after off-chain consensus. As can be seen from the figure, OpenChannel consumes more time than UpdateChannel. Moreover, combined with Fig. \ref{chain-time} (a), the service time is mainly used to process on-chain transactions.

\subsection{Overhead of CCBS}

The configuration of Xuperchain in OFBS is the same as \ref{exp_ofbs}. We simulated CCBS in Ubuntu 16.04 and shared the resources of 2.0GHZ 8-vCPUs, 16G, and Go1.16.4 linux/amd64. We exploited the homogeneous (FISCO BCOS and FISCO BCOS) and heterogeneous (Ethereum Ropsten and FISCO BCOS) blockchains to support seamless interactions. FISCO BCOS was also deployed in Ubuntu 16.04 and configured for four nodes and an organization. We connected Ethereum Ropsten through Infura. The number of relay nodes of CCBS is eight nodes if there are no special instructions. We compared with Swap \cite{tian2021enabling} to prove our efficiency since \textit{Swap} also took use of the proxy contract as a pivot. 

\begin{figure}[!t]
  \centering                                        
  \subfigure[Heto of CCBS]{            
  \includegraphics[width=1.5in]{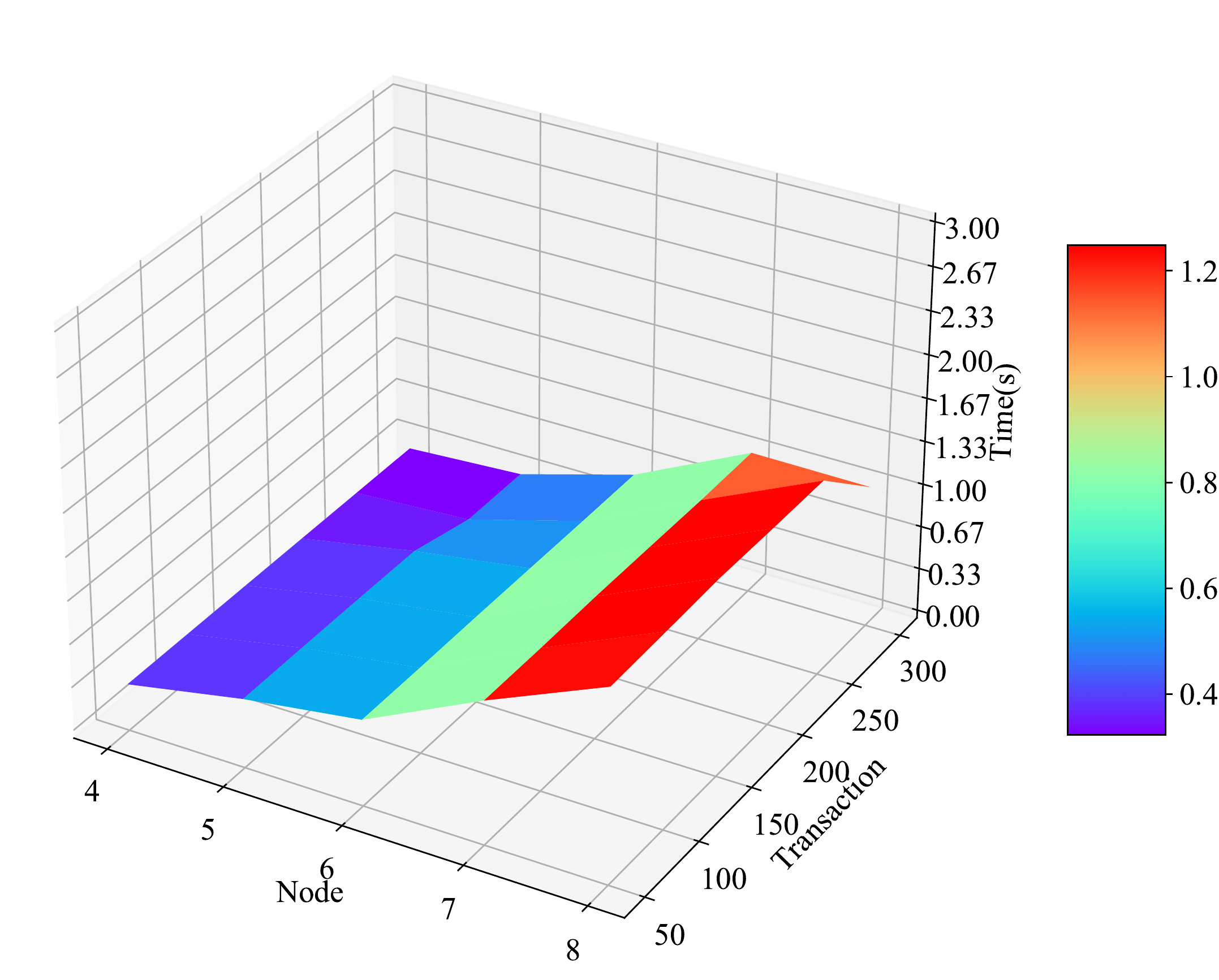}     
  }
  \subfigure[Heto of Swap]{
      \includegraphics[width=1.5in]{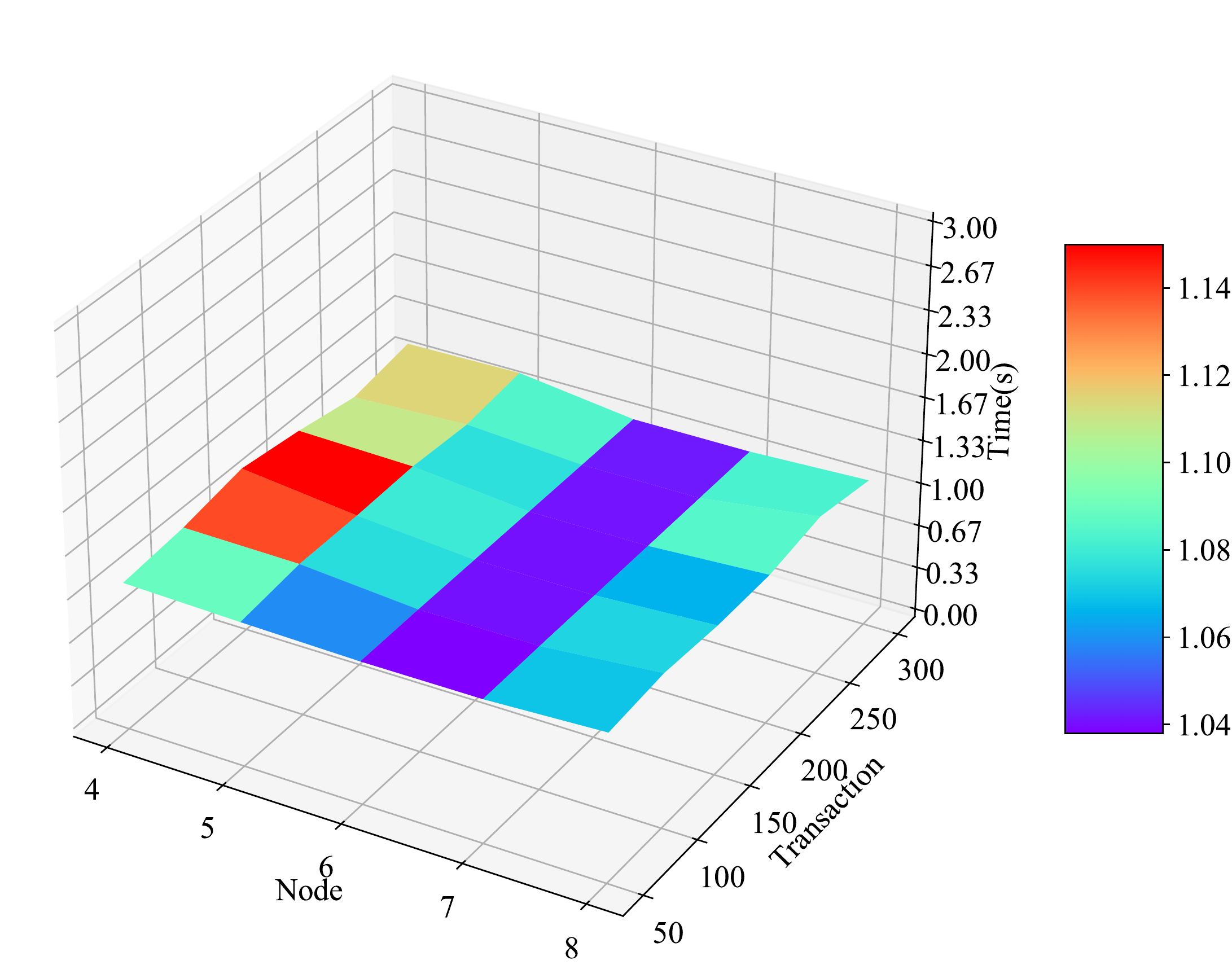}  
  }
  \subfigure[Homo of CCBS]{    
  \includegraphics[width=1.5in]{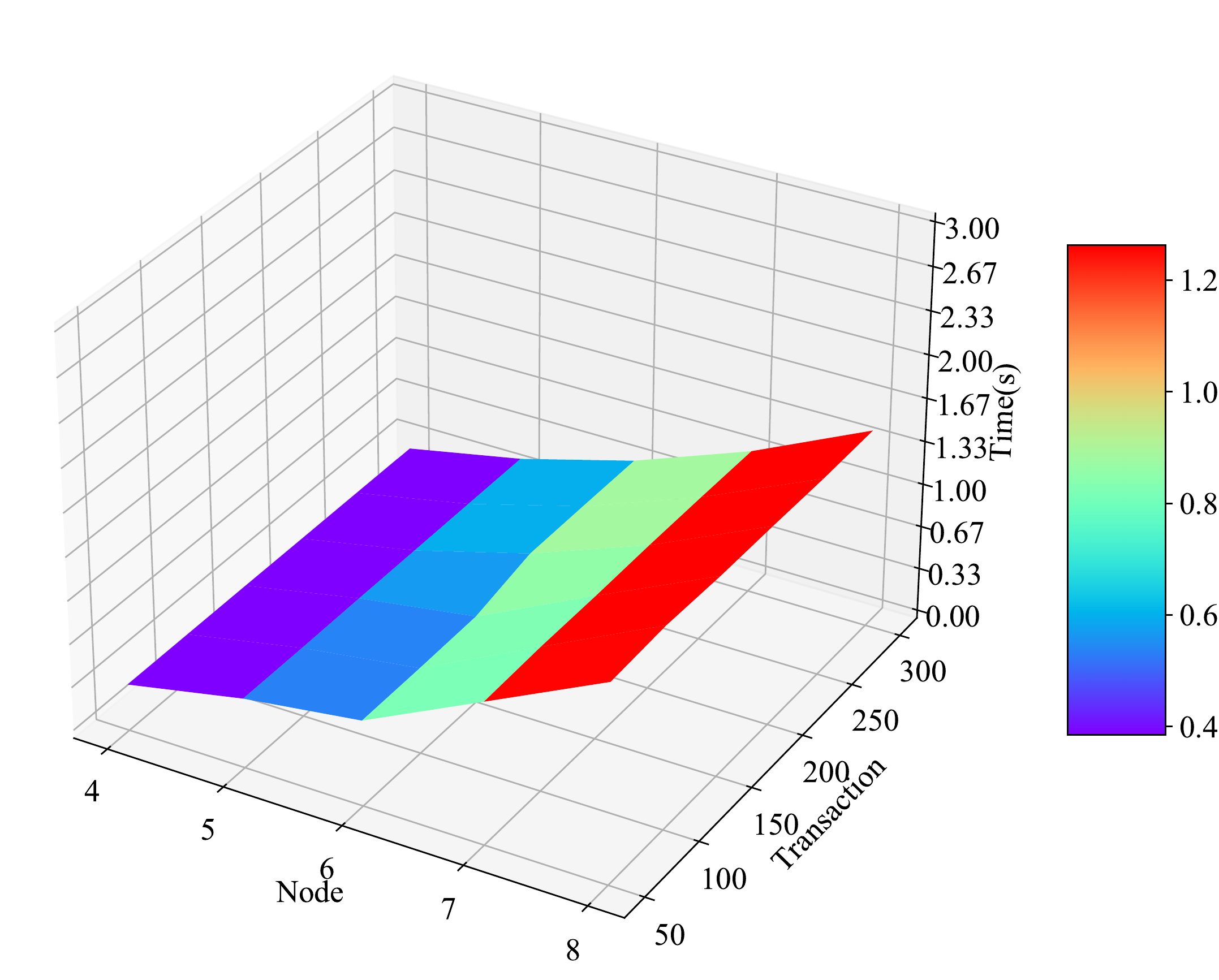}              
  }
  \subfigure[Homo of Swap]{
      \includegraphics[width=1.5in]{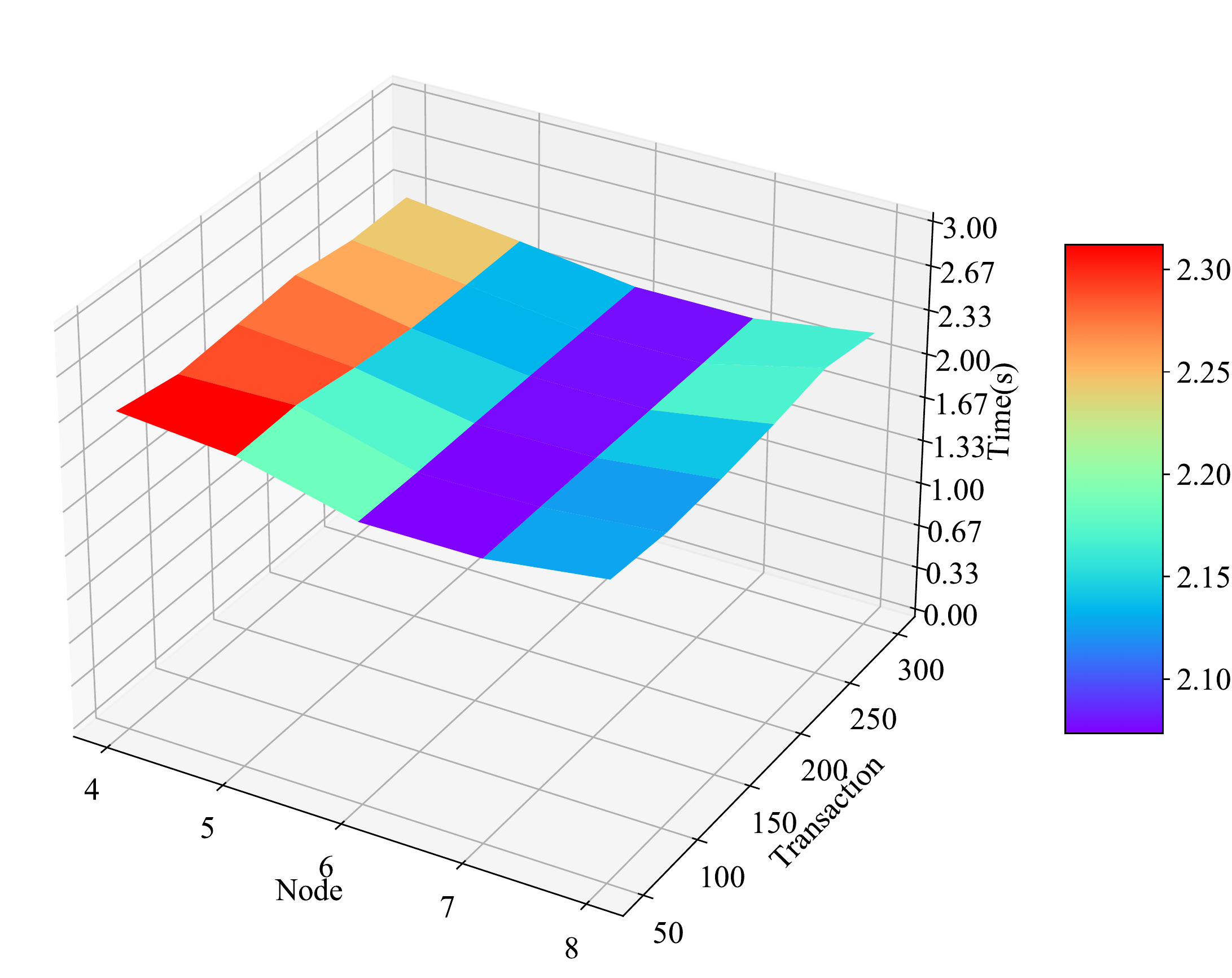} 
  }
  \caption{Consensus Time}                  
  \label{Cross consensus time} 
  \end{figure}

Fig. \ref{Cross consensus time} shows the off-chain consensus time when the number of consequent transactions [50, 300, 50] and relay nodes [4, 8, 1] increases. As can be seen from the figure, the off-chain consensus time of CCBS is reduced by about 30\% compared to Swap in homogeneous and heterogeneous groups. Moreover, the off-chain consensus time of CCBS in the two groups is almost the same, while that of Swap is relatively different. The reason is that CCBS has fewer interactions with blockchains than Swap.

\begin{figure}[!t]
  \centering                                                         
  \subfigure[Heto of CCBS]{            
  \includegraphics[width=1.5in]{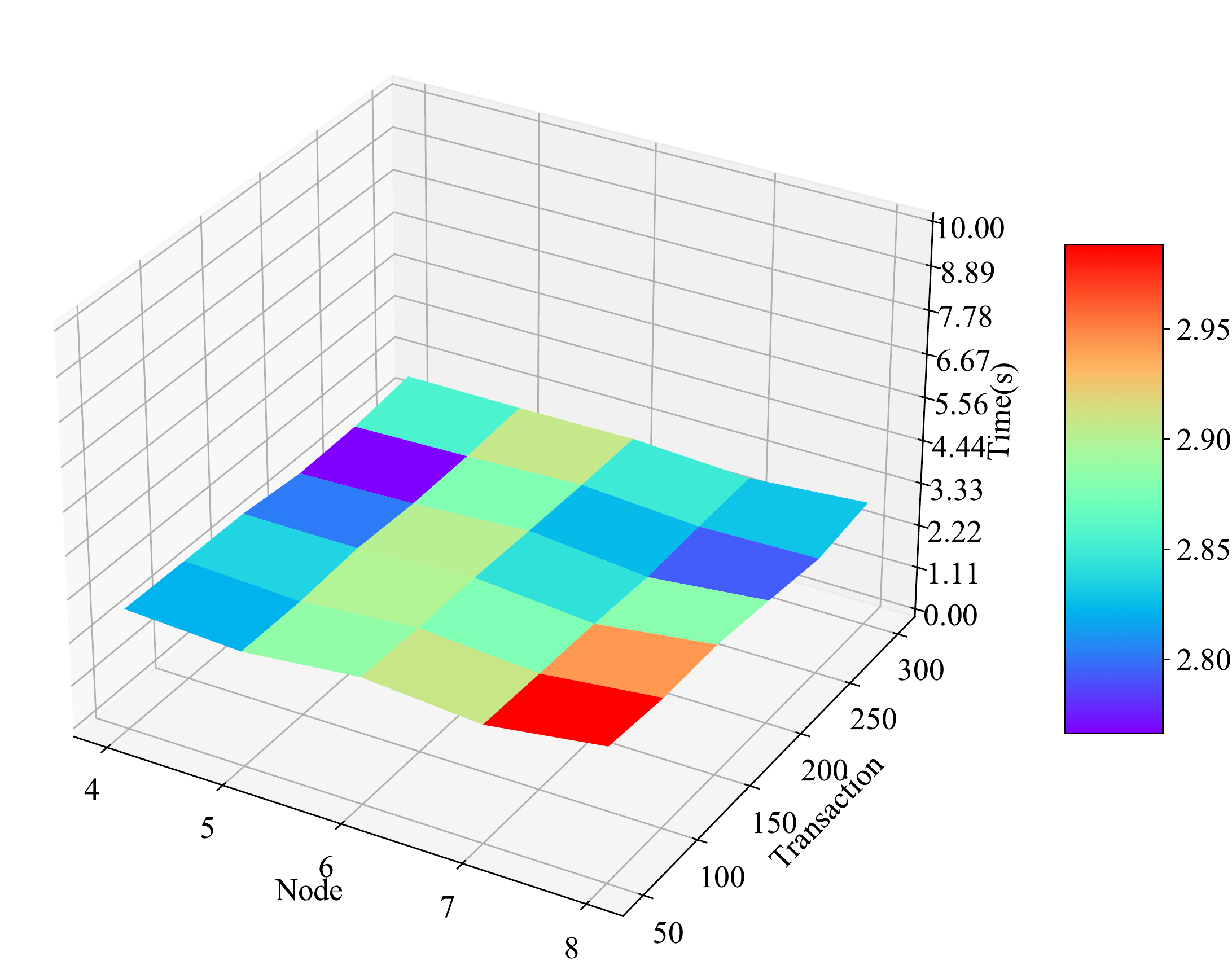}     
  }
  \subfigure[Heto of Swap]{
      \includegraphics[width=1.5in]{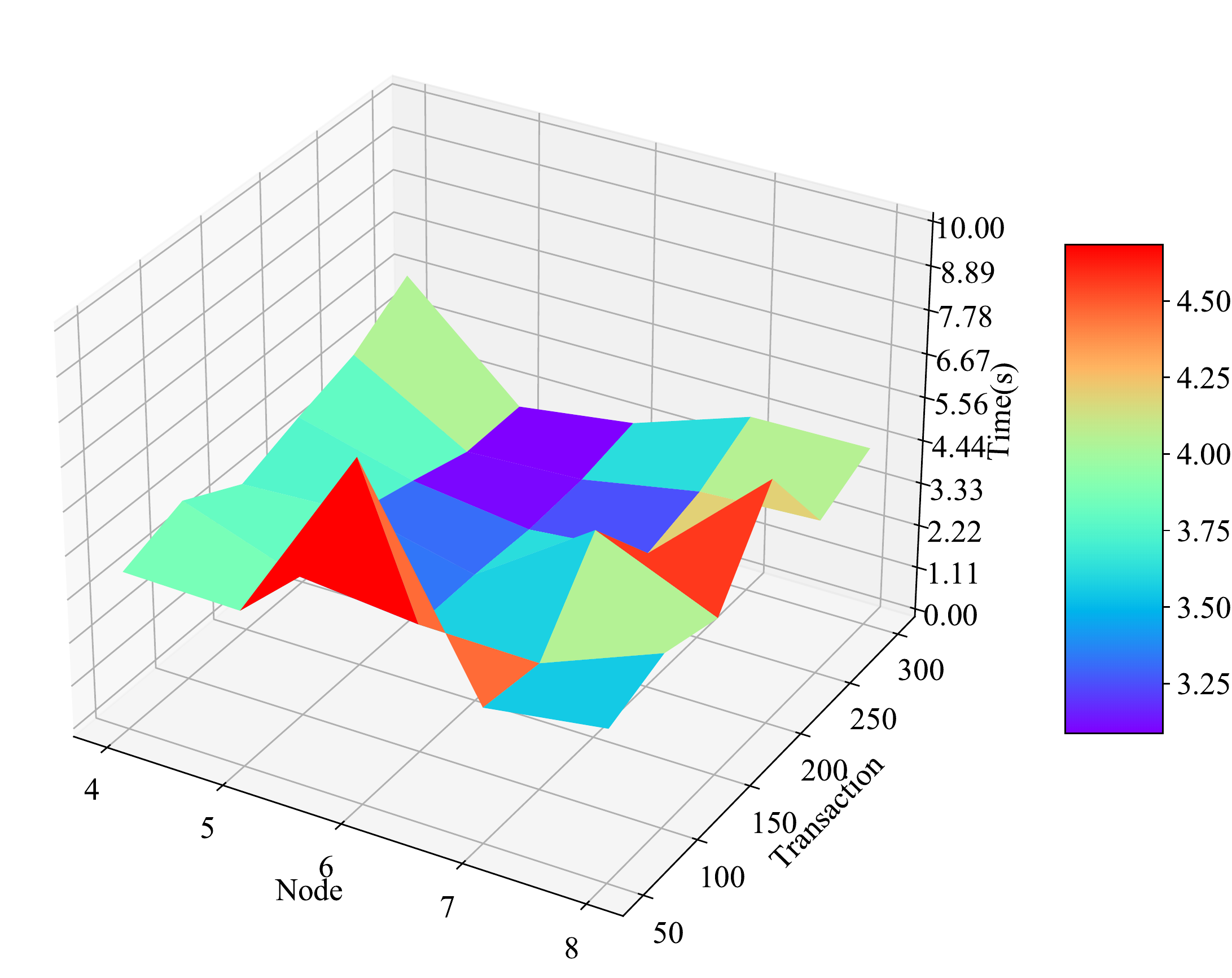}  
  }
  \subfigure[Homo of CCBS]{    
  \includegraphics[width=1.5in]{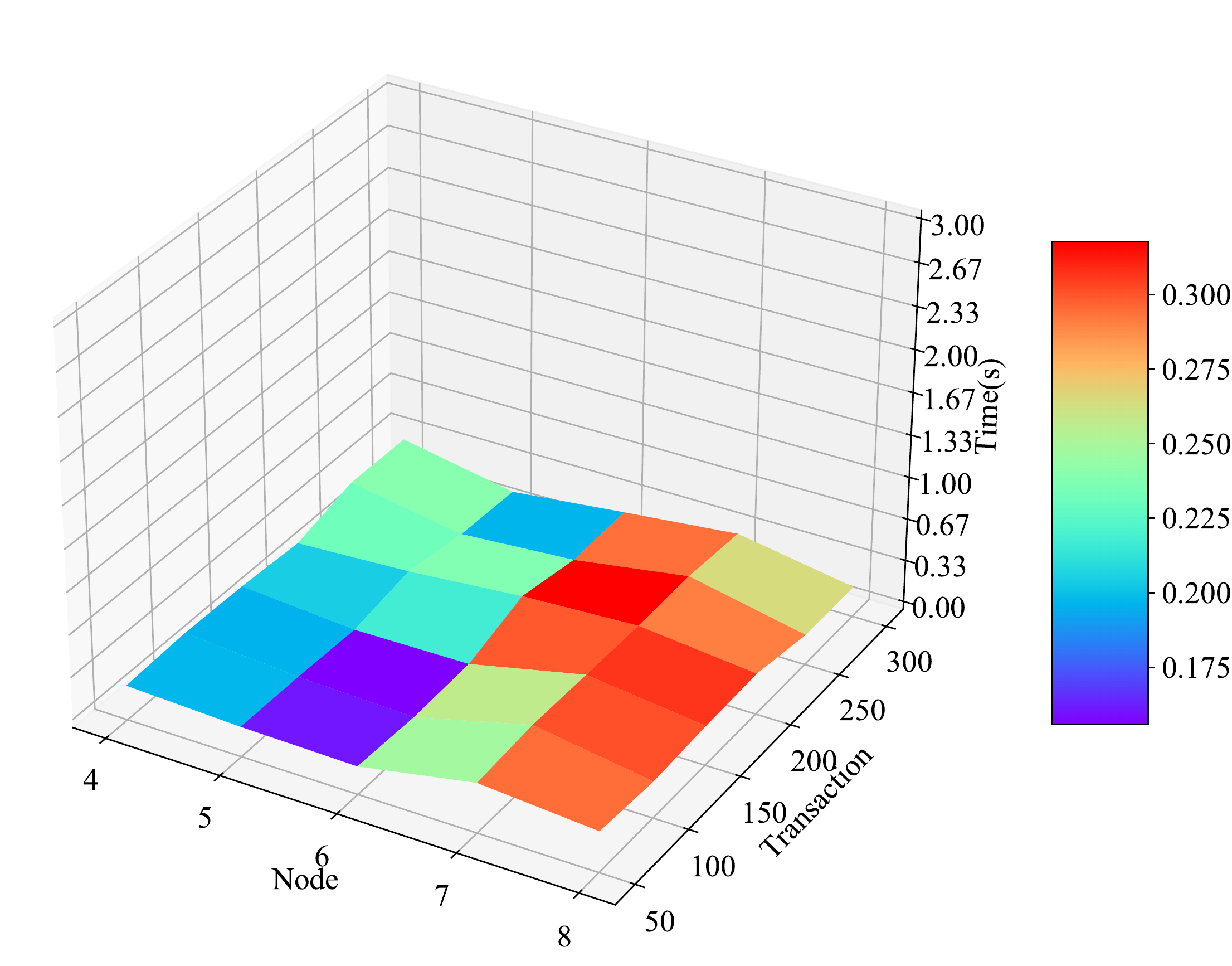}              
  }
  \subfigure[Homo of Swap]{
      \includegraphics[width=1.5in]{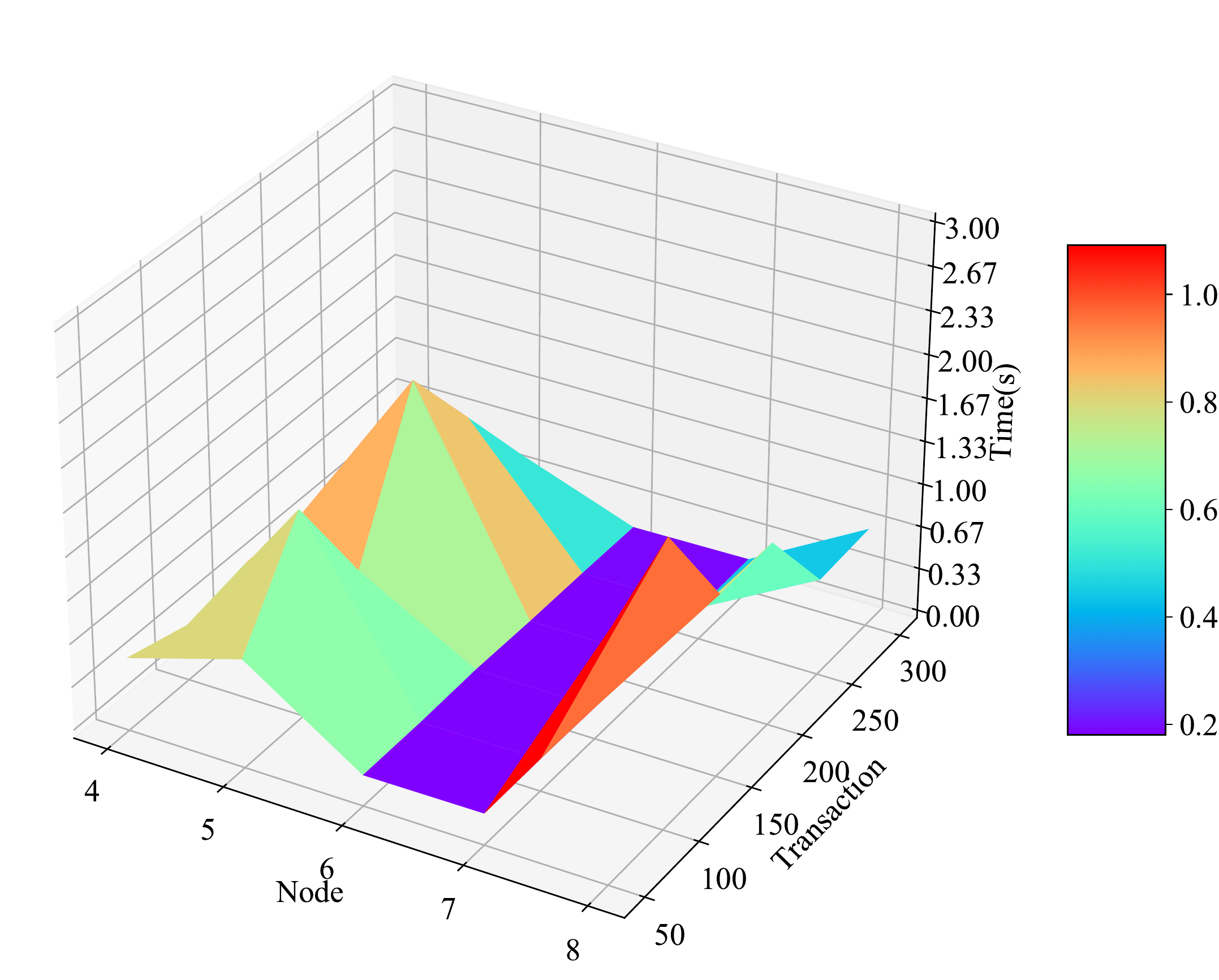} 
  }
  \caption{Validation Time}                  
  \label{Cross validate time} 
  \end{figure}

  Fig. \ref{Cross validate time} shows the validation time when the number of consequent transactions [50, 300, 50] and relay nodes [4, 8, 1] increases. As can be seen from the figure, the validation time of CCBS is reduced by 25\% compared to Swap in homogeneous and heterogeneous groups, which has the same reasons as above. Moreover, different approaches to deployment cause the time difference of CCBS since the homogeneous groups were deployed locally and part of heterogeneous groups were deployed by the cloud.
  
\begin{figure}[!t]
\centering   
    \subfigure[(Node, Concurrent)]{            
    \includegraphics[width=1in]{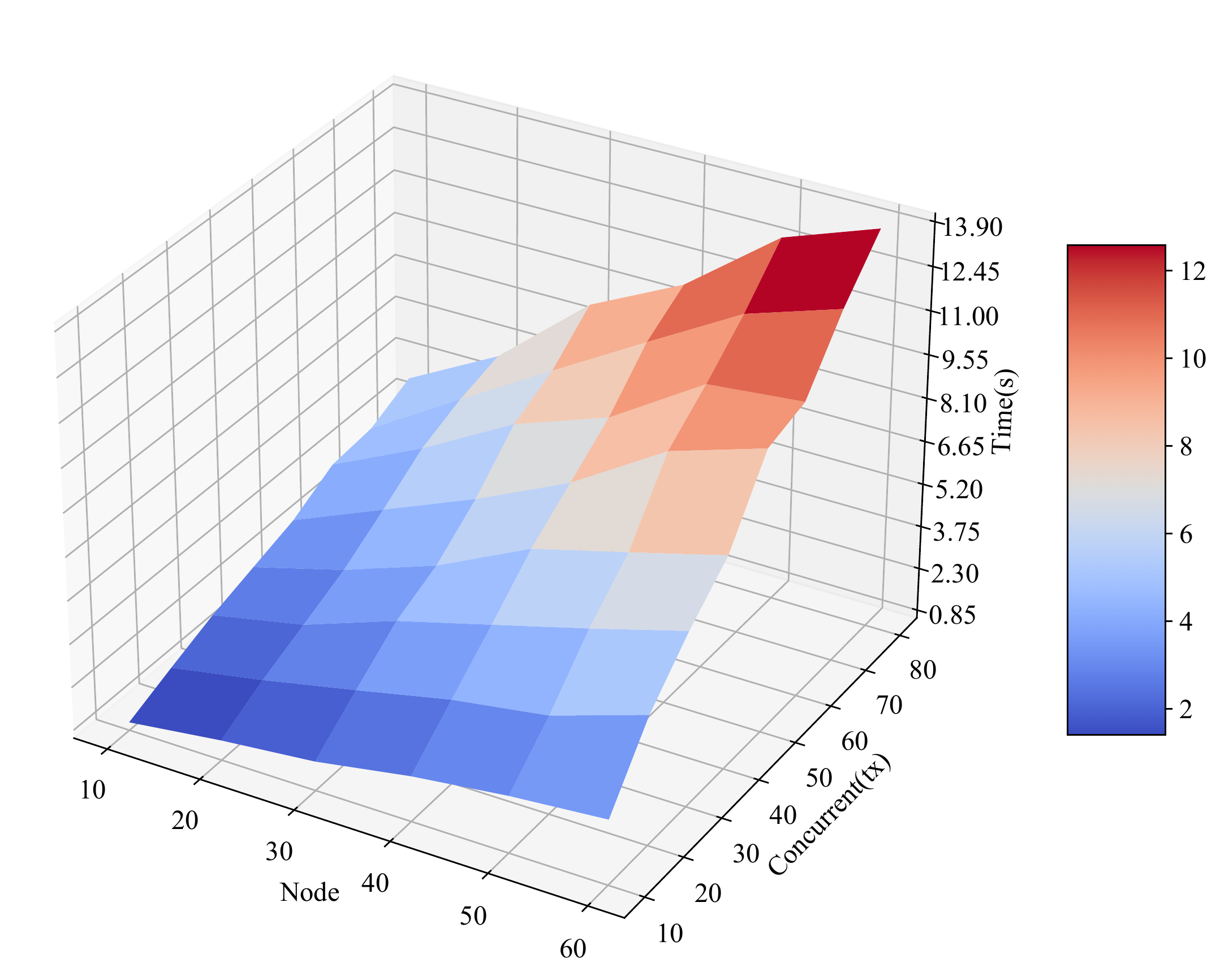}     
    }
    \subfigure[(Node, Size)]{    
    \includegraphics[width=1in]{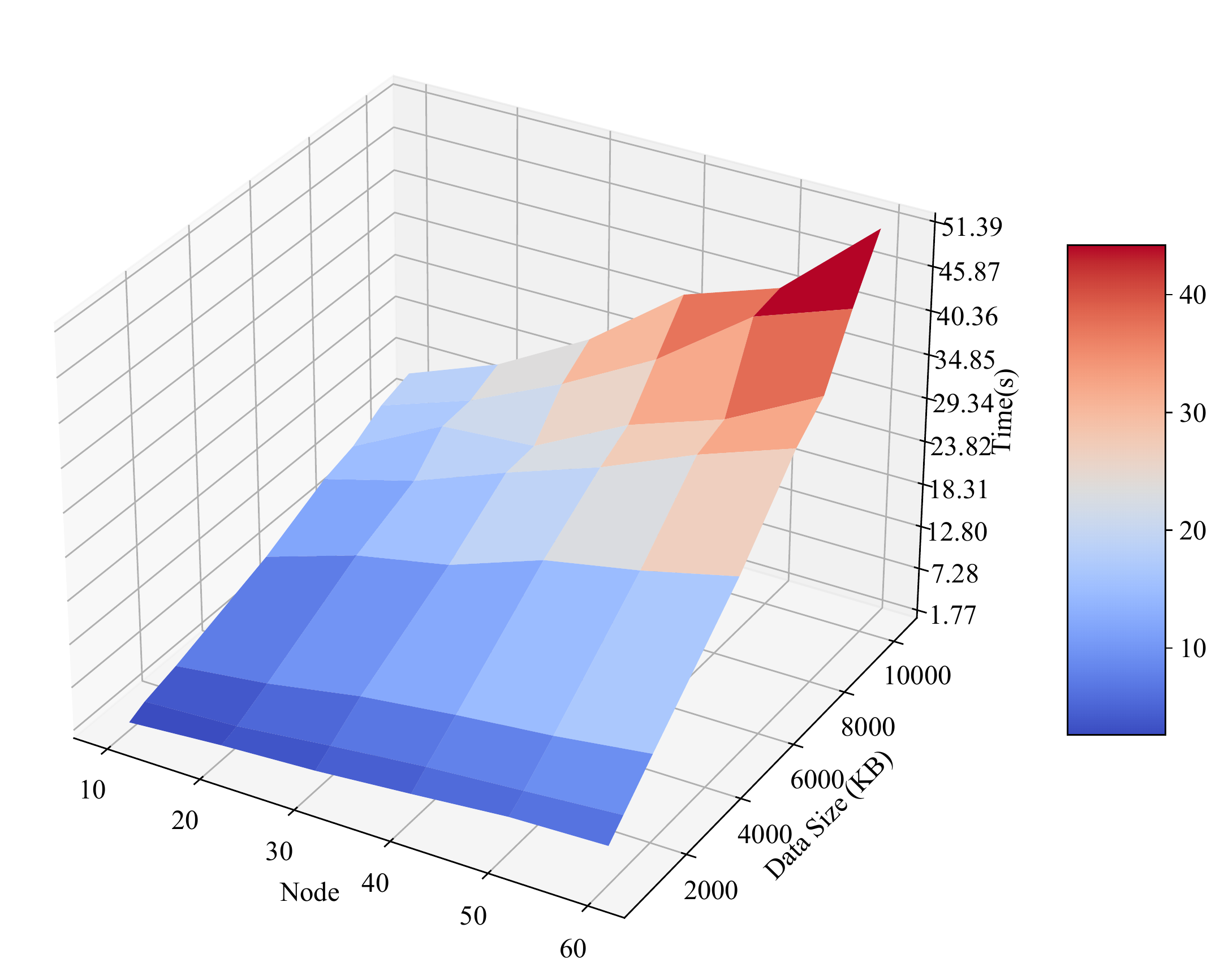}              
    }
    \subfigure[(Concurrent, Size)]{    
    \includegraphics[width=1in]{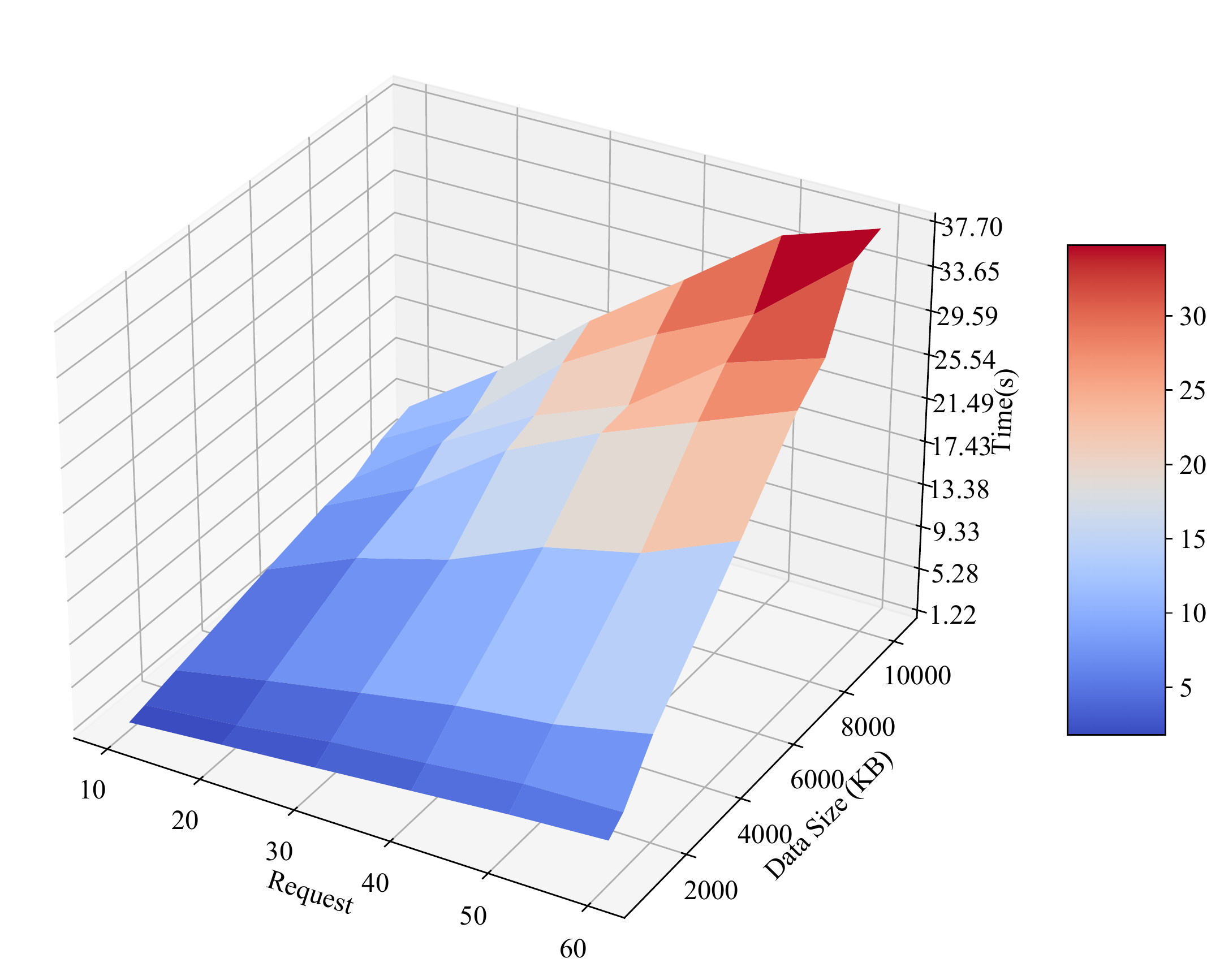}              
    }
    \caption{Process Time}
\label{Cross process time} 
\end{figure}

Fig. \ref{Cross process time} shows the performance of CCBS under large-scale networks when the concurrent transaction is  [10, 60, 10], the number of relay nodes is [10, 60, 10], and the data volumes are up to 10MB. It can be seen from the figure that CCBS has good performance in the large-scale network.

\subsection{Overhead of CPBS}

\begin{figure}[!t]
  \centering   
    \subfigure[Interaction]{            
    \includegraphics[width=3cm]{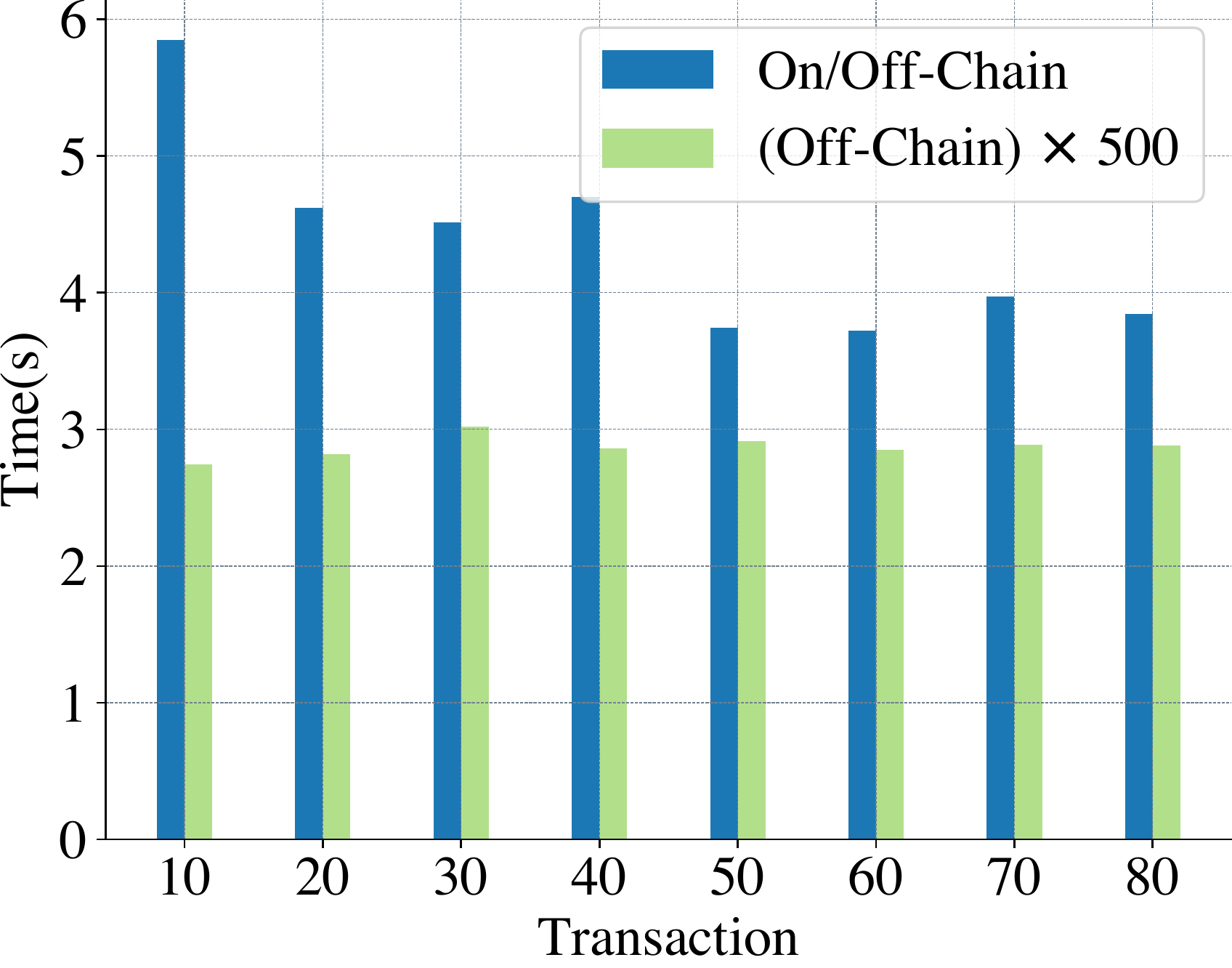}     
    }
    \subfigure[Data Size]{            
    \includegraphics[width=3cm]{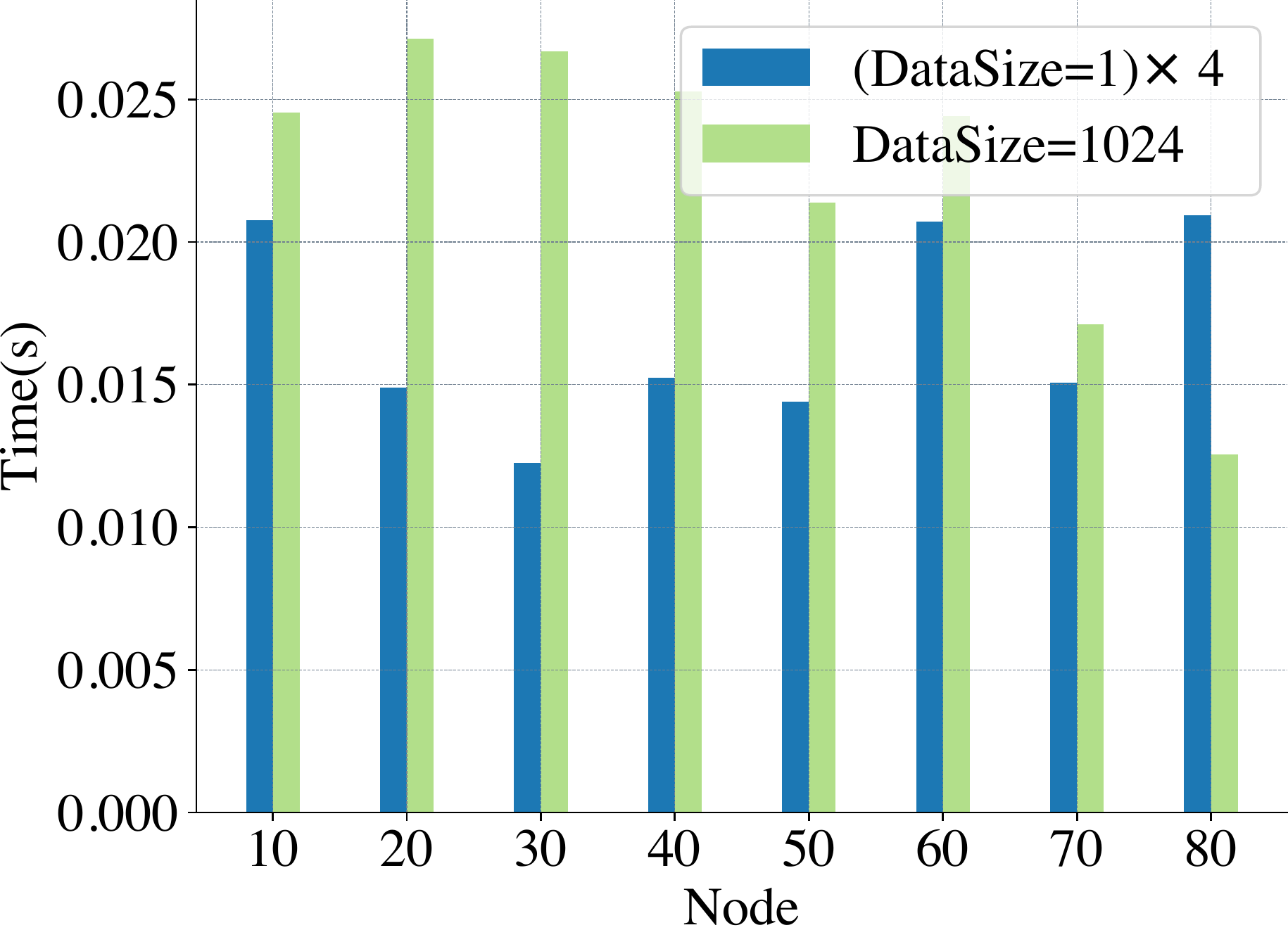}     }    
  \caption{Overhead}
  \label{Latency of ocbs} 
  \end{figure}

The configuration of Xuperchain in CPBS is the same as Section \ref{exp_ofbs}. Fig. \ref{Latency of ocbs} shows the overhead of CPBS when the number of consequent transactions and relay nodes are [10, 80, 10]. The blue in Fig. \ref{Latency of ocbs}(a) is the total overhead of the on-chain and off-chain interaction, and the green is the overhead of the off-chain interaction. Since the overhead of the off-chain interaction is too small, we amplify it by 500 times. It can be seen that the off-chain interaction of CPBS reflects the efficiency of CPBS. Fig. \ref{Latency of ocbs}(b) shows the overhead of CPBS when the data volume increases. It can be seen that the number of relay nodes has little effect on the latency of CPBS.

\section{Related Work}

Blockchain has derived much middleware to support different applications. This section briefly represents the most related state-of-art, including 1) blockchain middleware for offline networks, 2) cross-chain middleware for offline networks, and 3) computing Middleware for offline networks.

\subsection{Blockchain middleware for offline networks}

Yuntao W. \emph{et al.} \cite{wang2021disaster} proposed a lightweight blockchain-based collaborative framework for space-air-ground integrated network. They designed a delegated proof of stake consensus to share spare computing resources. Ming F. \emph{et al.} \cite{feng2019msnet} proposed a blockchain-based satellite communication network to protect safety. This approach can quickly detect and defend against cyber attacks. Chakrabarti C. \emph{et al.} \cite{chakrabarti2019blockchain} proposed a blockchain-based incentive scheme for the delay-tolerant network to establish emergency communication networks. Kongrath S. \emph{et al.} \cite{suankaewmanee2018performance} proposed MobiChain, a blockchain for mobile commerce. This mechanism connects blockchain with Sync Gateway \cite{ostrovsky2014synchronizing} via local direct connection or internet connection. However, those solutions are not work in offline networks.

Blockstream \cite{blockstream} exploited satellite to broadcast the Bitcoin blockchain around the world for free. SpaceChain \cite{spacechain}  is building a blockchain-based satellite network to deploy an Ethereum node. kryptoradio\cite{kryptoradio} exploit DVB-T broadcasting blockchain data, which is non-reliant on the internet. However, these solutions require specific equipment for receiving signals of the specific blockchain, which has expensive costs. And the above solutions can not process the cross-chain transaction and complex computation for offline clients in multiple blockchains.

\subsection{Cross-chain middleware for offline networks}

The mainstream cross-chain middleware includes Notary, Sidechain, Hash Locking, and Relay Chain \cite{buterin2016chain}.

Polkadot \cite{wood2016polkadot} proposed a relay chain-based cross-chain framework. Cosmos \cite{kwon2018network} proposed that all blockchains share Cross-chain Hub supported by Tendermint to complete data exchange. However, those solutions rely on a predefined system, which makes other blockchains need to adapt to this system. Moreover, not all blockchains can connect to the relay chain since they need to bid for slots.

Hyperledger Cactus \cite{cactus2020} proposed a blockchain integration solution based on hash locking\cite{hashlock2019}. The solution encapsulates a gateway layer and an interaction between validators and blockchains. Lys, L.  \textit{et al.} \cite{lys2020atomic} proposed an atomic cross-chain interaction scheme based on relayers and hash locking. Qi M. \textit{et al.} \cite{qi2020acctp} proposed a cross-chain transaction platform for high-value assets based on hash locking. However, those solutions are used for transferring assets between blockchains, which have limited applications. 

Jin H.  \textit{et al.} \cite{jin2018towards} proposed a passive cross-chain method based on monitor multiplexing reading. This method monitors the state of the network through a listener. Zhuotao L. \textit{et al.} \cite{liu2019hyperservice} proposed a secure interaction protocol for cross-chain transactions. Pillai, B. \textit{et al.} \cite{pillai2020cross} proposed a cross-chain interoperability protocol based on agents, which implements three-phase interactions between users and blockchains. However, those solutions can not consider malicious nodes in the protocol. Rui H. \cite{han2021vassago} proposed a cross-chain query method to implement an authentic provenance query. Tian H.  \textit{et al.} \cite{tian2021enabling} proposed a cross-chain asset transaction protocol based on validators and proxy contracts in Ethereum. However, this protocol elects validators based on proof of work, which leads to excessively long cross-chain transactions. The above solutions are all required to connect to the Internet, which is not suitable for offline clients.

\subsection{Computing Middleware for offline networks}

Harry K. \emph{et al.} \cite{kalodner2020blocksci} proposed BlockSci, a blockchain analysis platform based on an in-memory database. Weihui Y. \emph{et al.} \cite{yang2020ldv} proposed LDV, a method based on directed acyclic graph and historical data prune to reduce the storage overhead of blockchain. Xiaohai D. \emph{et al.} \cite{dai2020lvq} proposed LVQ, a lightweight verifiable query approach for Bitcoin, which is based on Bloom filter integrated Merkle Tree, Merkle Tree, and Sorted Merkle Tree. Haotian W. \emph{et al.} \cite{wu2021vql} proposes VQL, an efficient and verifiable cloud query service for blockchain. Cheng X. \emph{et al.} \cite{xu2018query} proposed APP, a data query structure of access-policy-preserving grid-tree based on Merkle Tree. Yijing L. \emph{et al.} \cite{lin2022novel} proposed a decentralized learning method based on oracles, which uses the data of off-chain producers to provide consumers with highly credible computing results. Muhammad M. \emph{et al.} \cite{muzammal2019renovating} proposed ChainSQL, a blockchain-based database system to achieve data modification of the blockchain and query speed of the distributed databases simultaneously. Saide Z. \emph{et al.} \cite{zhu2019zkcrowd} proposed zkCrowd, which distributes tasks through the public chain based on DpoS, and dynamically executes tasks through the private subchain based on PBFT. Weilin Z. \emph{et al.} \cite{zheng2019nutbaas} proposed NutBaaS, a blockchain-as-a-service platform that lowers development thresholds through log-based network real-time monitoring. However, the above solutions are all required to connect to the Internet, which is not suitable for offline clients.

\section{Conclusions and Future Works}

We introduce BcMON, including OFBS for offline clients accessing the blockchain, CCBS for offline clients accessing multiple blockchains, and CPBS for offline clients implementing complex on-chain computation. To the best of our knowledge, BcMON is the first blockchain middleware for offline networks. The prototype of BcMON has been implemented to evaluate the performance of the blockchain middleware. We will focus on modifying the network protocol to provide a more efficient offline network in future work.

\section*{ACKNOWLEDGEMENT}
This work is supported by National Natural Science Foundation of China (62072049), BUPT Excellent Ph.D. Students Foundation (CX2021133) and BUPT Innovation and Entrepreneurship Support Program (2022-YC-A112).

\bibliography{ref}
\bibliographystyle{IEEEtran}

\end{document}